\documentclass[showpacs,pra,aps,superscriptaddress,twocolumn,floatfix]{revtex4}

\usepackage{mathtools}
\usepackage[usenames,dvipsnames]{xcolor}
\usepackage{amsmath}
\usepackage{amssymb,mathrsfs}
\usepackage{graphicx}

\usepackage{mathtext}

\newcommand{\prt}{\partial}
\newcommand{\la}{\lambda}

\newcommand{\al}{\alpha}
\newcommand{\om}{\omega}

\newcommand{\sn}{\mathrm{sn}}
\newcommand{\cn}{\mathrm{cn}}

\newcommand{\red}{\color{red}}

\begin{document}

\title{Riemann problem for the photon fluid: self-steepening effects}

\author{S. K. Ivanov} \affiliation{Institute of Spectroscopy, Russian
Academy of Sciences, Troitsk, Moscow, 108840, Russia}
\affiliation{Moscow Institute of Physics and Technology, Institutsky
lane 9, Dolgoprudny, Moscow region, 141700, Russia}

\author{A. M. Kamchatnov}
\affiliation{Institute of Spectroscopy, Russian Academy
of Sciences, Troitsk, Moscow, 108840, Russia}
\affiliation{Moscow Institute of Physics and Technology, Institutsky
lane 9, Dolgoprudny, Moscow region, 141700, Russia}

\begin{abstract}
We consider the Riemann problem of evolution of initial discontinuities for the photon fluid
propagating in a normal dispersion fiber with account of self-steepening effects. The dynamics
of light field is described by the nonlinear Schr\"odinger (NLS) equation with self-steepening term
appearing due to retardation of the fiber material response to variations of the electromagnetic
signal. It is shown that evolution dynamics in this case is much richer than that for the NLS
equation. Complete classification of possible wave structures is given for all possible jump
conditions at the discontinuity.
\end{abstract}

\pacs{42.65.Tg, 42.81.Dp, 47.37.+q, 02.30.Ik}

\maketitle



\section{Introduction}\label{sec1}

Dispersive shock waves (DSWs, or undular bores), that are oscillatory wave structures emerging in evolution
of wave pulses after wave breaking, are ubiquitous, being observed in various physical systems including
water waves, Bose-Einstein condensates, waves in magnetics and in nonlinear optics (see, e.g., review article
\cite{eh-2016} and references therein).
In nonlinear optics, the formation of temporal dispersive shock waves was observed in single-mode
optical fibers for the wavelength corresponding to a normal group velocity dispersion upon steepening
of powerful picosecond optical pulses acquiring almost rectangular shapes and linear frequency
chirp due to combined action of the self-phase-modulation and dispersion effects \cite{tsj-1985}.
Formation of dispersive shock wave in \cite{tsj-1985} was identified by means of analysis of the spectrum
of transmitted pulses, while in the subsequent work \cite{rg-1989} the evidence of the shock wave formation
was obtained already in the time domain. Optical shock waves were observed not only in light pulses,
but also in light beams. For example, the propagation of high-intensity localized beams superimposed
on low-intensity plane-wave background led to formation of both one- and two-dimensional
spatial shock waves in photo-refractive crystals with defocusing nonlinearity \cite{wjf-2007}
and allowed observation of interactions between several shocks. A shock fan filled with
non-interacting one-dimensional gray solitons that emanate from a gradient catastrophe
developing around the notch of powerful dark beam in defocusing optical medium was observed
in Ref.~\cite{conti-2009}. Recently, a fiber-optics analogue of the dam-breaking phenomenon
was studied experimentally in Ref.~\cite{xu-2017}.

Theoretically, the DSWs are represented as modulated nonlinear periodic waves and then the process
of their formation and evolution is described by the Whitham theory of modulations (for a review see
Ref.~\cite{eh-2016}). In the fiber optics applications, the dynamics of pulses is described usually
by the nonlinear Schr\"odinger (NLS) equation that accounts for two main effects---quadratic normal dispersion
and Kerr nonlinearity. For this case, the theory of DSWs is already well developed and the main parameters
of the arising wave structures can be calculated for typical idealized situations in simple
analytical form. In particular, consideration of many realistic problems can be reduced to
analysis of the so-called Riemann problem of evolution of discontinuity in the initial data.
Such a discontinuity can appear, for example, as a jump in the time dependence of the
light intensity, what is most typical in physics of light pulses in fibers, or evolve from
a ``collision'' of two pulses in which case not only intensity has a discontinuity but
also the time and space derivatives of the phase. Classification of possible wave structures
in the NLS equation theory was given in Refs.~\cite{gk-1987,el-1995}, and it provides the theoretical
basis for calculation of characteristic parameters of such experiments as that of Ref.~\cite{xu-2017}.

However, in nonlinear optics, besides quadratic dispersion and Kerr nonlinearity, many other
effects can play important role in propagation of pulses. For example, in experiment \cite{wjf-2007}
with photo-refractive material the saturation of nonlinearity is quite essential and the
corresponding theory of DSWs was developed in Ref.~\cite{el-2007}. In fiber optics, one
needs to take into account such effects as dissipation, higher-order dispersion, intra-pulse
Raman scattering and self-steepening (see, e.g., \cite{ka-2003}). These effects can drastically
change evolution of DSWs leading sometimes to violation of the supposition that such an evolution is adiabatically
slow, as apparently it happens in the case of considerable higher order dispersion \cite{cbt-2014}.
On the other hand, small dissipation can stop spreading out of the oscillatory region so that its
width is stabilized with the size being inverse proportional to the dissipation coefficient. These
effects have been studied in different physical contexts and their role in nonlinear optics
seems to be quite clear. The self-steepening effects are usually
described by the last term in the {\it modified NLS} ({\it mNLS}) equation which can be written in
non-dimensional form as
\begin{equation}\label{DNLS1}
    \begin{split}
    { i}q_x+\frac{1}{2}q_{tt}\pm |q|^2q-{ i}\al\left(|q|^2q\right)_t=0.
    \end{split}
\end{equation}
In early publication \cite{al-1983} it was shown in dispersionless approximation that during
an evolution the pulse acquires an asymmetric form instead of gradual symmetric deformation of
its form in the NLS equation theory. This observation demonstrates the most unusual feature
of the self-steepening term caused by retardation of dielectric response in optical fibers,
namely, lack of time inversion symmetry: the equation (\ref{DNLS1}) is not invariant with respect
transformation $t\mapsto-t,q\mapsto q^*$. To reach the initial form of the equations, one needs to make an
additional inversion transformation $x\mapsto -x$. This means that the ``right'' and ``left''
directions are not equivalent to each other, that is the flow of `optical fluid' is anisotropic.
Inclusion of dispersion can stabilize the self-steepening
wave breaking resulting in the soliton mode of pulse propagations and the corresponding
soliton solutions of the so-called ``derivative nonlinear Schr\"{o}dinger equation'' (DNLS equation)
\begin{equation}\label{dnls0}
  iq_x+\frac12q_{tt}-i(|q|^2q)_t=0
\end{equation}
related with (\ref{DNLS1}) were found in \cite{kn-1978}, its multi-soliton solution in \cite{vch-1989},
and periodic solutions in \cite{kamch-1990}. However, the role of the self-steepening term in
evolution of DSWs has not get the full solution so far. The DNLS equation (\ref{dnls0}) appears also in the
theory of nonlinear Alfv\'en waves in magnetized plasma (see, e.g., \cite{kbhp-1988,hkb-1989,mjolhus-1989}),
but again only part of possible wave structures appearing after wave breaking were studied by the Whitham
method in Ref.~\cite{gke-1992}.

The aim of this paper is to give full solution to the problem of evolution of an initial
discontinuity in framework of the Whitham approach to the mNLS equation (\ref{DNLS1}). Although the
Whitham equations that govern slow evolution of modulated periodic waves in this case
were derived already in Ref.~\cite{kamch-1990}, their application to this problem is not trivial
because of non-standard properties of the dispersionless equations that do not satisfy the
so-called condition of {\it genuine nonlinearity} and the method of Refs.~\cite{gp-1973}
(KdV equation) and \cite{el-1995} (NLS equation) is not applied directly. In simpler situation of
unidirectional waves whose evolution is governed by the modified KdV (or Gardner) equation,
this problem was solved in Ref.~\cite{kamch-2012} where it was found that in addition to
DSWs and rarefaction waves the arising structures can also include trigonometric and
combined shocks or kinks depending on the sign of the higher order nonlinearity (see also
earlier papers \cite{marchant-2008,ep-2011} where partial similar results were also obtained).
In this paper we extend this method to the equation (\ref{DNLS1}) describing evolution of
nonlinear pulses in fibers.

The paper is structured as follows. In section \ref{sec2} we consider the linear waves propagation along a
constant background with the aim to derive the corresponding dispersive relations for two
different wave modes and to
illustrate the above mentioned lack of time inversion in the pulse propagation. The weakly
nonlinear waves are discussed in section \ref{sec3} where we show that in weakly nonlinear case these
two modes obey either to KdV or mKdV equation what results in very different their behavior.
In section \ref{sec4} we obtain the periodic solutions to equation (\ref{DNLS1}) by the finite-gap integration
method which yields these solutions in the form convenient for applications in the Whitham
theory of modulations and the Whitham equations are also derived in section \ref{sec4}. In section \ref{sec5}
we describe the elementary wave structures that appear as building blocks in the general
wave patterns. In section \ref{sec6} we apply the developed
theory to derivation of the full classification of wave structures arising in evolution of
the initial discontinuities. The last section \ref{sec7} is devoted to conclusions.

\section{Linear waves}\label{sec2}

\begin{figure}[t] \centering
\includegraphics[width=8cm]{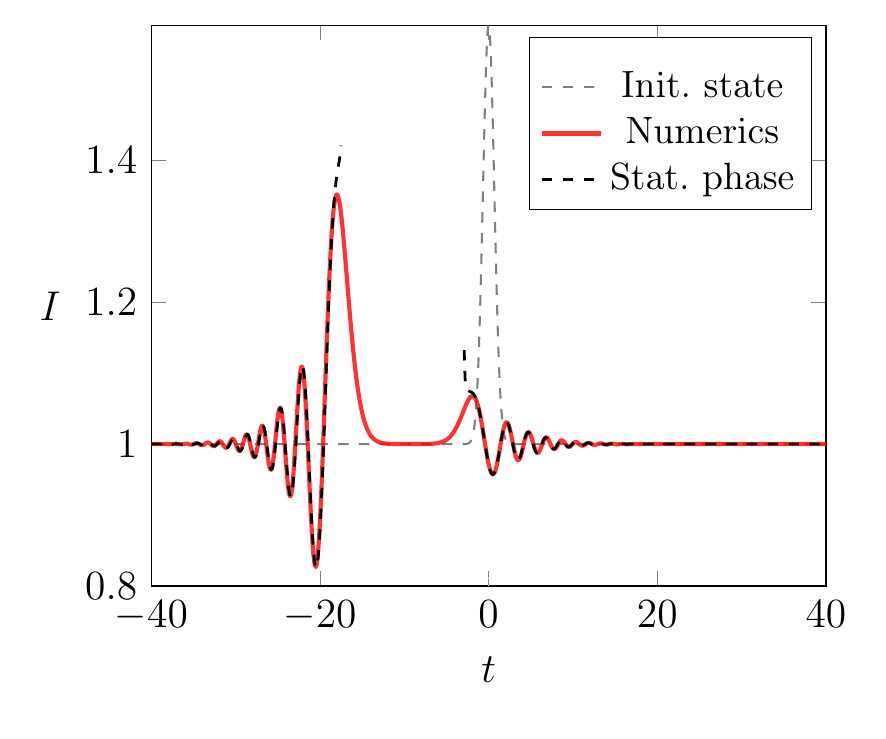}
\caption{Evolution of a pulse in the linear approximation for the mNLS equation (\ref{DNLS1}) with $\al=1$.
        Dashed thin line shows the initial disturbance, thick dashed lines correspond to the stationary phase
        approximation (\ref{StPhase}) and red (solid gray) line to numerical evaluation of the solution (\ref{IntForPer}) at $x=5$
        for the initial disturbance (\ref{InitialDsit}) with $a=0.95$.}
        \label{Fig1}
\end{figure}

We shall start with the study of linear waves in a waveguide along a uniform wave background
with the amplitude $\sqrt{I_0}=|q_0|=\mathrm{const}$. It is more convenient to make a substitution
$q(t,x)=\tilde{q}(t,x)\exp{(-{ i}I_0 x)}$ and then the  modified NLS equation (\ref{DNLS1})
with ``minus'' sign in the Kerr nonlinearity (normal dispersion) and $\alpha>0$
transforms to
\begin{equation}\label{DNLS2}
    \begin{split}
    { i}\tilde{q}_x+\frac{1}{2}\tilde{q}_{tt}+\left(I_0-|\tilde{q}|^2\right)\tilde{q}
    -{ i} \al\left(|\tilde{q}|^2\tilde{q}\right)_t=0.
    \end{split}
\end{equation}
We suppose that at the undisturbed state the phase is everywhere equal to zero and
linearize the equation with respect to small disturbance
\begin{equation}\label{}
    \begin{split}
    \tilde{q}=\sqrt{I_0}+\delta q,\qquad |\delta q|\ll\sqrt{I_0}
    \end{split}
\end{equation}
to obtain equation for $\delta q$:
\begin{equation}\label{DNLSLin}
    \begin{split}
    { i} \delta {q}_x+\frac{1}{2}\delta {q}_{tt}-I_0(\delta q
    +\delta q^*)-{ i}\al I_0 (2\delta {q}_t+\delta {q}^*_t)=0.
    \end{split}
\end{equation}
This equation should be solved with the initial condition $\delta {q}|_{x=0}=\delta {q_0}(t)$.
After separation of the real and imaginary parts
$    \delta {q}=A+{ i}B,$
we obtain the system from which we can exclude $B$ and get the linear equation
\begin{equation}\label{EqForA}
    \begin{split}
    A_{xx}+I_0\left(3\al^2I_0-1\right)A_{tt}+\frac{1}{4}A_{tttt}-4\al I_0A_{xt}=0.
    \end{split}
\end{equation}
It can be readily solved by the Fourier method. To this
end, we note that linear harmonic waves $A\propto\exp{[{ i}(k x-\omega t)]}$
satisfy to the dispersion law
\begin{equation}\label{disp-law}
    \begin{split}
    k_{1,2}(\om)=\om \left( -2\al I_0\pm \sqrt{\frac{\om^2}{4}+I_0(\al^2I_0+1)} \right).
    \end{split}
\end{equation}
After standard calculations we arrive at the solution expressed in terms of the
Fourier transform $I_0(\om)$ of the initial (input) intensity disturbance $ I'(x,t)=2\sqrt{I_0}A(x,t)$,
\begin{equation}\label{IntForPer}
    \begin{split}
     I'(t,x) & = J_1(x,t)-J_2(x,t),\\
    J_1 & = \int_{-\infty}^{\infty} \delta I(\om) K_1(\om) e^{{ {i}} xf_1(\om)} \frac{d\om}{2\pi}, \\
    J_2 & = \int_{-\infty}^{\infty} \delta I(\om) K_2(\om) e^{{ i} xf_2(\om)} \frac{d\om}{2\pi},
    \end{split}
\end{equation}
where
\begin{equation}\label{}
    \begin{split}
     K_1(\om)=\frac{k_1(\om)+3\al I_0\om}{k_2(\om)-k_1(\om)},\quad
     K_2(\om)=\frac{k_2(\om)+3\al I_0\om}{k_2(\om)-k_1(\om)}, \quad
    \end{split}
\end{equation}
and
\begin{equation}\label{}
    \begin{split}
    f_1(\om)=k_1(\om)-\om \frac{t}{x}, \quad
    f_2(\om)=k_2(\om)-\om \frac{t}{x}.
    \end{split}
\end{equation}
These integrals can be estimated for large distance of propagation $x$ by the method of stationary phase
resulting in
\begin{equation}\label{StPhase}
    \begin{split}
    J_1 & \simeq\frac{2\delta I_0(\om^{(1)}_0)K_2(\om^{(1)}_0)}{\sqrt{2\pi x
    {|\frac{d^2f_1}{d\om^2}|}_{\om_0^{(1)}}}} \cos{ \left (xf_1(\om^{(1)}_0)+\frac{\pi}{4} \right )}, \\
    J_2 & \simeq\frac{2\delta I_0(\om^{(2)}_0)K_1(\om^{(2)}_0)}{\sqrt{2\pi x
    {|\frac{d^2f_2}{d\om^2}|}_{\om_0^{(2)}}}} \cos{ \left (xf_2(\om^{(2)}_0)+\frac{\pi}{4} \right )},
    \end{split}
\end{equation}
where $\om^{(1)}_0$ and $\om^{(2)}_0$ are the values of $\om$ at the
points of the stationary phase that are defined by the equations
\begin{equation}\label{}
    \begin{split}
    \frac{df_1}{d\om}=0,\qquad \frac{df_2}{d\om}=0.
    \end{split}
\end{equation}
In Fig.~\ref{Fig1} we compare the numerical calculation of the integral (\ref{IntForPer}) with its approximate
estimation (\ref{StPhase}) for the initial perturbation
\begin{equation}\label{InitialDsit}
    \begin{split}
     I'(t)=\frac{1}{\sqrt{\pi}a}\exp{\left(-\frac{t^2}{a^2}\right)},\quad
     I'(\om)=\exp{\left(-\frac{\om^2a^2}{4}\right)}.
    \end{split}
\end{equation}
As we see, the pulse splits into two smaller pulses, however, on the contrary to the NLS case, they are
not symmetrical pulses propagating in opposite directions. Now these two pulses have different profiles
and propagate with different group velocities. This is manifestation of lack of the time inversion invariance
mentioned in the introduction, which is caused by the last term in the mNLS equation (\ref{DNLS1}).
It should be noted that the asymptotic solution (\ref{StPhase})
describes well the wave packet even for not very large $x$.

The two modes of propagation in the linear approximation to the equation (\ref{DNLS1}) differ from each other
not only by the dispersive properties of their propagation; in fact, their nonlinear properties are also
drastically different, as we shall see in the next section.

\section{Small dispersion and weak nonlinear limits of the modified NLS equation}\label{sec3}

We are interested in the leading dispersive and nonlinear corrections to the dispersionless linear propagation of
disturbances along the background pulse. Therefore they can be considered separately and
after that their contributions should be added to give the resulting approximate equation.

In the small dispersion limit the series expansion of the expressions (\ref{disp-law}) in
degrees of $\om$ yields
\begin{equation}\nonumber
\begin{split}
k(\om)\cong& \om\Bigg\{-2\al I_0 \\
&\pm\left[\sqrt{I_0(1+\al^2I_0)}+\frac{\om^2}{8\sqrt{I_0(1+\al^2I_0)}}\right]\Bigg\}.
\end{split}
\end{equation}
This approximation of dispersion laws corresponds to linear equations for propagation of, say,
small disturbances of intensity, $I=I_0+I'$,
\begin{equation}\label{295.23}
\begin{split}
  \frac{\prt I'}{\prt x}& +\left[\pm\sqrt{I_0(1+\al^2I_0)}-2\al I_0\right]\frac{\prt I'}{\prt t}\\
  & \mp\frac{1}{8\sqrt{I_0(1+\al^2I_0)}}\frac{\prt^3 I'}{\prt t^3}=0.
  \end{split}
\end{equation}

To find small nonlinear corrections, we turn to the dispersionless limit which can be
obtained by means of well-known Madelung transformation
\begin{equation}\nonumber
    \begin{split}
    q(t,x)=\sqrt{I(t,x)}\exp{\left({ i}\int^tu(t',x)dt'\right)}
    \end{split}
\end{equation}
which after substitution into the mNLS equation (\ref{DNLS1}) and separation of real
and imaginary parts yields the system
\begin{equation}\label{Hydro}
    \begin{split}
    & I_x+\left(uI-\frac{3}{2}\al I^2\right)_t=0, \\
    & u_x+uu_t+I_t-\al(uI)_t+\left(\frac{I^2_t}{8I^2}-\frac{I_{tt}}{4I}\right)_t=0.
    \end{split}
\end{equation}
The last term in the second equation describes the dispersion effects and the full system
will be considered later. Now we shall discuss the dispersionless limit when this term is omitted
and we arrive at the hydrodynamic system
\begin{equation}\label{Hydro2}
    I_x+\left(uI-\frac{3}{2}\al I^2\right)_t=0, \quad
    u_x+uu_t+I_t-\al(uI)_t=0,
\end{equation}
where the first equation can be interpreted as the continuity equation for the intensity
$I$ and the second one as the Euler equation for the ``flow velocity'' $u$. This system can
be cast in standard way to the Riemann diagonal form
\begin{equation}\label{211.5}
  \frac{\prt r_{\pm}}{\prt x}+\frac1{v_{\pm}}\frac{\prt r_{\pm}}{\prt t}=0
\end{equation}
for the Riemann invariants
\begin{equation}\label{RiemannInv}
 {r_{\pm}}=\frac{u}2-\al I\pm\sqrt{I(1+\al^2I-\al u)}
\end{equation}
with inverse velocities
\begin{equation}\label{211.4b}
  \frac1{v_{\pm}}=u-2\al I\pm\sqrt{I(1+\al^2I-\al u)}.
\end{equation}
As one can see, if we put chirp $u$ equal to zero, then we reproduce the low frequency
limit $\om\to0$ of the inverse phase velocities $k/\om$ of linear waves given by
Eq.~(\ref{disp-law}), as it should be. This means that the two linear modes correspond
to the linear approximation of the so-called simple waves with one of the Riemann constant.
Hence, the weakly nonlinear waves correspond to the next order approximation of these
simple waves with respect to amplitude of propagating disturbance. Since properties of
these two modes are very different, they should be considered separately.

\subsection{Korteweg-de Vries mode}

In dispersionless approximation, the KdV equation is obtained in the case of
the weakly nonlinear simple wave
evolution with constant Riemann invariant $r_+=\mathrm{const}$. Assuming that a pulse propagates
along the same background $I=I_0$, $u=0$, we have the relation between $I$ and $u$,
$$
\frac{u}2-\al I+\sqrt{I(1+\al^2I-\al u)}=-\al I_0+\sqrt{I(1+\al^2I_0)},
$$
which defines $u$ as a function of $I$, $u=u(I)$, along this simple wave. This function can be
substituted into the varying Riemann invariant $r_-$ and the corresponding inverse velocity $1/v_-$.
Hence, series expansion of equation (\ref{211.5}) for $r_-$ with respect to small disturbance $I'$
of the intensity $I=I_0+I'$ yields the weakly nonlinear approximation
\begin{equation}\nonumber
  \begin{split}
  \frac{\prt I'}{\prt x}&-(\sqrt{I_0(1+\al^2I_0)}+2\al I_0)\frac{\prt I'}{\prt t}\\
  &-\frac32(\sqrt{1+\al^2I_0} +\sqrt{\al^2I_0})I'\frac{\prt I'}{\prt t}=0.
  \end{split}
\end{equation}
Combining the nonlinear term from this equation and the small dispersion effects described by equation (\ref{295.23}),
we arrive at the KdV equation
\begin{equation}\label{297.2}
\begin{split}
\frac{\prt I'}{\prt x}&-(\sqrt{I_0(1+\al^2I_0)}+2\al I_0)\frac{\prt I'}{\prt t}\\
&-\frac{3(\sqrt{1+\al^2I_0} +\sqrt{\al^2I_0})}{2\sqrt{I_0}}I'\frac{\prt I'}{\prt t}\\
&+\frac{1}{8\sqrt{I_0(1+\al^2I_0)}}\frac{\prt^3 I'}{\prt t^3}=0.
\end{split}
\end{equation}
It is worth noticing that in both limits $\al^2I_0\gg1$ and $\al^2I_0\ll1$ the nonlinear term
has finite value and we need not to include higher order corrections for
taking into account higher order nonlinear effects. The situation is different for another simple wave
in the small amplitude approximation.

\subsection{Gardner mode}

Derivation of evolution equation for another weakly nonlinear simple wave is similar,
however, at it will be clear from the result, now we have to take into account the terms of the second
order approximation. In this case the Riemann invariant $r_-=\mathrm{const}$ is constant and $u=u(I)$
is defined now by the relation
$$
\frac{u}2-\al I-\sqrt{I(1+\al^2I-\al u)}=-\al I_0-\sqrt{I(1+\al^2I_0)}.
$$
Substitution of this $u=u(I)$ into $r_+$, $v_+$ and series expansion with respect to $I'$
up to the second degree of $I'$ as well as taking into account the dispersion effects
according to equation (\ref{295.23}) gives the Gardner equation
\begin{equation}\label{297.5}
\begin{split}
\frac{\prt I'}{\prt x}&+(\sqrt{I_0(1+\al^2I_0)}-2\al I_0)\frac{\prt I'}{\prt t}\\
&+\frac{3(\sqrt{1+\al^2I_0} -\sqrt{\al^2I_0})}{2\sqrt{I_0}}I'\frac{\prt I'}{\prt t}\\
&-\frac{3(\sqrt{1+\al^2I_0} +\sqrt{\al^2I_0})}{8I_0\sqrt{I_0}}I^{\prime 2}\frac{\prt I'}{\prt t}\\
&-\frac{1}{8\sqrt{I_0(1+\al^2I_0)}}\frac{\prt^3 I'}{\prt t^3}=0.
\end{split}
\end{equation}
In the limit $\al^2I_0\gg1$ this equation reduces to the mKdV equation
\begin{equation}\label{297.6}
  \frac{\prt I'}{\prt x}-\al I_0\frac{\prt I'}{\prt t}-\frac{3\al}{4I_0}I^{\prime 2}\frac{\prt I'}{\prt t}
  -\frac1{8\sqrt{I_0(1+\al^2I_0)}}\frac{\prt^3 I'}{\prt t^3}=0.
\end{equation}
Disappearance of quadratic nonlinearity from this equation was the reason why the cubic
nonlinearity was included into equation (\ref{297.5}). In the opposite limit $\al^2I_0\ll1$, neglecting
higher order correction, we return to the KdV equation
\begin{equation}\label{297.7}
   \frac{\prt I'}{\prt x}+\sqrt{I_0} \frac{\prt I'}{\prt t}+\frac3{2\sqrt{I_0}}I' \frac{\prt I'}{\prt t}
   -\frac1{8I_0}\frac{\prt^3 I'}{\prt t^3}=0.
\end{equation}
It differs from analogous limit of equation (\ref{297.2}) by the replacement $t\to-t$, that is
in this limit the symmetry between the left and right propagating waves is restored.

Formation of DSWs from initial discontinuities in the KdV equation theory is well known since
the pioneering paper Ref.~\cite{gp-1973} --- the initial discontinuity evolves into either
rarefaction wave or cnoidal DSW. However, situation for the mKdV equation is much more
complicated \cite{kamch-2012} and in this case we can get eight different structures
including, besides the rarefaction waves and cnoidal DSWs, also trigonometric DSWs,
combined shocks and their combinations separated by plateau. Therefore one should expect
that in the case of Riemann problem for the equation (\ref{DNLS1}) we have also to get much richer
structure than in the NLS case. To solve this problem, at first we have to find periodic
solutions of the equation (\ref{DNLS1}) in convenient for us form, that is in the form parameterized by the
parameters related with the Riemann invariants of the corresponding Whitham modulation
equations by simple formulae. This is achieved by the restricted finite-gap integration
method developed in Ref.~\cite{kamch-1990}, and in the next section we shall obtain the
periodic solutions by this method and derive the Whitham equations.

\section{Periodic solutions and Whitham equations}\label{sec4}

The finite-gap integration method (see, e.g., \cite{kamch-2000})  is based on possibility
of representing of the mNLS equation (\ref{DNLS1}) as a compatibility condition of two
systems of linear equations with a spectral parameter $\la$
\begin{align}
            \label{lax1}
    &\frac{\partial}{\partial t}
    \begin{pmatrix}
        {\psi}_1 \\
        {\psi}_2 \\
    \end{pmatrix}
    =\left(
    \begin{array}{cc}
        {F} & {G} \\
        {H} & -{F} \\
    \end{array}
    \right)
    \begin{pmatrix}
        {\psi}_1 \\
        {\psi}_2 \\
    \end{pmatrix}\;, \\
            \label{lax2}
    &\frac{\partial}{\partial x}
    \begin{pmatrix}
        {\psi}_1 \\
        {\psi}_2 \\
    \end{pmatrix}
    =\left(
    \begin{array}{cc}
        {A} & {B} \\
        {C} & -{A} \\
    \end{array}
    \right)
    \begin{pmatrix}
        {\psi}_1 \\
        {\psi}_2 \\
        \end{pmatrix}\;,
\end{align}
where
\begin{equation}\label{}
    \begin{split}
    & {F}=-2{ i}\left(\lambda^2-\frac{1}{4\al}\right),\quad {G}=2\sqrt{\al}\,\lambda q,\quad {H}=2\sqrt{\al}\,\lambda q^*, \\
    & {A}=-{ i}\left\{4\left(\lambda^2-\frac{1}{4\al}\right)^2+2\al\lambda^2|q|^2\right\}, \\
    & {B}=\sqrt{\al}\left\{4\lambda\left(\lambda^2-\frac{1}{4\al}\right) q+\lambda\left({ i} q_t+2\al|q|^2q\right)\right\}, \\
    & {C}=\sqrt{\al}\left\{4\lambda\left(\lambda^2-\frac{1}{4\al}\right) q^*-\lambda\left({ i} q^*_t-2\al|q|^2q^*\right)\right\}.
    \end{split}
\end{equation}
This Lax pair can be obtained by simple transformation from the known Lax
pair for the DNLS equation (\ref{dnls0}) (see Ref.~\cite{WadatiKI-79}).
Here $2\times2$ linear problems (\ref{lax1}) and (\ref{lax2}) have two linearly
independent basis solutions which we denote as ($\psi_1$, $\psi_2)^T$
and ($\varphi_1$, $\varphi_2)^T$. We define ``squared basis function'' by the formulae
\begin{equation}\label{}
    \begin{split}
    {f}=-\frac{ i}2(\psi_1\varphi_2+\psi_2\varphi_1), \quad
    {g}=\psi_1\varphi_1,          \quad
    {h}=-\psi_2\varphi_2.
    \end{split}
\end{equation}
They obey the linear equations
\begin{subequations}\label{fght}
    \begin{align}
    {f}_t=& { i}G h-{ i}H g, \label{ft} \\
    {g}_t=& 2F g+2{ i}G f, \label{gt} \\
    {h}_t=& -2F h-2{ i}H f, \label{ht}
    \end{align}
\end{subequations}
and
\begin{subequations}\label{fghx}
    \begin{align}
    {f}_x=& { i}B h-{ i}C g, \label{fx} \\
    {g}_x=& 2A g+2{ i}B f, \label{gx} \\
    {h}_x=& -2A h-2{ i}C f. \label{hx}
    \end{align}
\end{subequations}
We look for the solutions of these equations in the form
\begin{equation}\label{sol:fgh}
    \begin{split}
    {f} & = \left(\lambda^2-\frac{1}{4\al}\right)^2 - {f}_1 \left(\lambda^2-\frac{1}{4\al}\right) + {f}_2, \\
    {g} & = \sqrt{\al}\, \lambda \left(\lambda^2-\frac{1}{4\al}-\frac{\mu}{2}\right) q, \\
    {h} & = \sqrt{\al}\, \lambda \left(\lambda^2-\frac{1}{4\al}-\frac{\mu^*}{2}\right) q^*.
    \end{split}
\end{equation}
Here the functions ${f}_1(x,t)$, ${f}_2(x,t)$, $\mu(x,t)$ and $\mu^*(x,t)$
are unknown;  $\mu(x,t)$ and $\mu^*(x,t)$ are not interrelated {\it a priori}, but we
shall find soon that they are complex conjugate, whence the notation.

It is easy to check that the expression ${f}^2-{g}{h} = P(\lambda)$ is independent of $x$ and $t$,
and periodic solutions are distinguished by the condition that $P(\la)$ be a polynomial in $\la$ in accordance
with the ansatz (\ref{sol:fgh}),
\begin{equation}\label{}
    \begin{split}
    {f}^2-{g}{h} & = P(\lambda) = \prod_{i=1}^4\left(\lambda^2-\lambda_i^2\right)= \\
    & = \lambda^8-s_1 \lambda^6+s_2 \lambda^4-s_3\lambda^2+s_4.
    \end{split}
\end{equation}
Equating the coefficients of like powers of
$\lambda$ at two sides of this identity, we get
\begin{subequations}\label{s1234}
    \begin{align}
    s_1 & = \frac{1}{\al}+2{f}_1+\al|q|^2, \label{s1} \\
    s_2 & = {f}_1^2+\frac{3}{2\al}{f}_1+2{f}_2+\frac{3}{8\al^2}+ \frac{1}{2}|q|^2+\frac{1}{2}\al|q|^2(\mu+\mu^*),
    \label{s2} \\
    \begin{split}
    s_3 & = \frac{{f}_1^2}{2\al}+\frac{3{f}_1}{8\al^2}+2f_1{f}_2+\frac{{f}_2}{\al}+\frac{1}{16\al^3}+\frac{|q|^2}{16\al}+ \\
    & \quad + \frac{1}{8}|q|^2(\mu+\mu^*)+\frac{1}{4}\al|q|^2\mu\mu^*,
    \end{split}
    \label{s3} \\
    s_4 & = \left(\frac{f_1}{4\al}+f_2\right)^2+\frac{1}{8\al^2}\left(\frac{f_1}{4\al}+{f}_2\right)+\frac{1}{256\al^4}.
    \label{s4}
    \end{align}
\end{subequations}
Here $s_i$ are standard symmetric functions of the four
zeros $\lambda_i^2$ of the polynomial,
\begin{equation}\label{}
    \begin{split}
    & s_1=\sum_i\lambda_i^2,\quad s_2=\sum_{i<j}\lambda_i^2\lambda_j^2,
    \quad s_3=\sum_{i<j<k}\lambda_i^2\lambda_j^2\lambda_k^2, \\
    & s_4=\lambda_1^2\lambda_2^2\lambda_3^2\lambda_4^2.
    \end{split}
\end{equation}

Equations (\ref{s1234}) allow us to express $\mu,\mu^*$ as functions of $I=|q|^2$.
The last equation (\ref{s4}) gives
\begin{equation}\label{f1f2}
    \begin{split}
    {f}_1 & = \frac{1}{2}\left(s_1-\frac{1}{\al}-\al I\right), \\
    {f}_2 & = \frac{1}{8\al^2}\left(1+\al^2I-\al s_1\pm8\al^2\sqrt{s_4}\right).
    \end{split}
\end{equation}
We substitute that into (\ref{s2}) and (\ref{s3}) and obtain
the system for $\mu$ and $\mu^*$, which can be easily solved to give
\begin{equation}\label{muR}
    \begin{split}
    \mu =& \frac{1}{4\al I}\Big[4s_2-\left(s_1-\al I\right)^2-2I\\
    & \pm8\sqrt{s_4}+{ i}\sqrt{-\mathcal{R}\left(\al I\right)}\Big],
    \end{split}
\end{equation}
where
\begin{equation}\label{tildeR}
    \begin{split}
    \mathcal{R}(\nu)= & \nu^4-4s_1\nu^3+\left(6s_1^2-8s_2\pm48\sqrt{s_4}\right)\nu^2- \\
                & - \left(4s_1^3-16s_1s_2+64s_3\pm32s_1\sqrt{s_4}\right)\nu\\
                &+\left(-s_1^2+4s_2\pm8\sqrt{s_4}\right)^2.
    \end{split}
\end{equation}
The introduced here function $\mathcal{R}$ is a fourth-degree polynomial in $\nu$
and it is called an {\it algebraic resolvent} of the polynomial $P(\lambda)$,
because zeros of $\mathcal{R}(\nu)$  are
related to zeros of $P(\lambda)$ by the following simple symmetric
expressions: the upper sign ($+$) in (\ref{tildeR}) corresponds to the
zeros
\begin{equation}\label{zeros1}
    \begin{split}
    \nu_1 & =(-\lambda_1+\lambda_2+\lambda_3+\lambda_4)^2, \\
    \nu_2 & =(\lambda_1-\lambda_2+\lambda_3+\lambda_4)^2, \\
    \nu_3 & =(\lambda_1+\lambda_2-\lambda_3+\lambda_4)^2, \\
    \nu_4 & =(\lambda_1+\lambda_2+\lambda_3-\lambda_4)^2,
    \end{split}
\end{equation}
and the lower sign ($-$) in equation (\ref{tildeR}) corresponds to the zeros
\begin{equation}\label{zeros2}
    \begin{split}
    \nu_1 & =(-\lambda_1+\lambda_2+\lambda_3-\lambda_4)^2, \\
    \nu_2 & =(\lambda_1-\lambda_2+\lambda_3-\lambda_4)^2, \\
    \nu_3 & =(\lambda_1+\lambda_2-\lambda_3-\lambda_4)^2, \\
    \nu_4 & =(\lambda_1+\lambda_2+\lambda_3+\lambda_4)^2. \\
    \end{split}
\end{equation}
This can be proved by a simple check of the Viet\'e formulae.

Substitution of Eqs.~(\ref{sol:fgh}) into Eqs.~(\ref{fght})
gives after equating the coefficients of like powers of $\lambda$
expressions for the time derivatives of ${f}_1$ and ${f}_2$
\begin{equation}\label{f1t:f2t}
    \begin{split}
    f_{1,t} = { i}\al|q|^2(\mu-\mu^*), \quad f_{2,t}=-\frac{1}{4\al}f_{1,t},
    \end{split}
\end{equation}
and of $q$ and $\mu$
\begin{equation}\label{}
    \begin{split}
    q_t=2{ i}q\left({\mu}-2{f}_1\right), \quad (\mu q)_t=-8{ i}q{f}_2.
    \end{split}
\end{equation}
In a similar way, substitution of (\ref{sol:fgh}) into (\ref{fghx}) with account of (\ref{f1t:f2t})
gives equations for the space derivatives of ${f}_1$ and ${f}_2$
\begin{equation}\label{eq42n}
    \begin{split}
    f_{1,x} = \left(2f_1+\alpha|q|^2\right)f_{1,t}, \quad f_{2,x}=-\frac{1}{4\alpha}f_{1,x}.
    \end{split}
\end{equation}
As follows from (\ref{s1}), the first equation (\ref{eq42n}) gives the expression for the constant phase velocity
\begin{equation}\label{}
    \begin{split}
    \frac{1}{V}=-(2{f}_1+\alpha|q|^2)=\frac{1}{\al}-s_1,
    \end{split}
\end{equation}
and $f_1$ depends on $\xi=t-x/V$ only. Then from the first equation (\ref{f1f2}) we see that
the intensity $I$ also depends only on $\xi$. The equations for dynamics of $I$ can be easily
found by substitution of (\ref{muR}) into the first equation (\ref{f1t:f2t}) with account
again of equation (\ref{f1f2}), so we get
\begin{equation}\label{}
    \begin{split}
    \frac{d(\al I)}{d\xi}=\sqrt{-\mathcal{R}(\al I)},
    \end{split}
\end{equation}
where $\mathcal{R}$ is, as we know, a fourth degree polynomial with the zeros given in terms of $\la_i$
by the formulae (\ref{zeros1}) or (\ref{zeros2}). This equation can be solved in standard way in terms of
elliptic functions. Without going to much detail we shall present here the main results.

We shall assume that $\la_i$ are ordered according to $\la_1\leq \la_2\leq\la_3\leq\la_4<0$
and then both our definitions (\ref{zeros1}) and (\ref{zeros2}) give the same ordering
of $\nu_i$: $\nu_1\leq\nu_2\leq\nu_3\leq\nu_4$. The inverse phase velocity can be
written as
\begin{equation}\label{phaseV}
    \begin{split}
    \frac{1}{V}=\frac{1}{\al}-\sum_{i=1}^4\la^2_i=\frac{1}{\al}-\frac14\sum_{i=1}^4\nu_i.
    \end{split}
\end{equation}
The real solutions correspond to oscillations of $\al I$ within the intervals where
$-\mathcal{R}(\al I)\geq0$.

(A) At first we shall consider the periodic solution corresponding to oscillations
of $\al I$ in the interval
\begin{equation}\label{eq18}
    \nu_1\leq \al I\leq \nu_2.
\end{equation}
Standard calculation yields, after some algebra, the solution in terms of Jacobi
elliptic functions:
\begin{equation}\label{eq20}
    \al I=\nu_2-\frac{(\nu_2-\nu_1)\cn^2(\theta,m)}{1+\frac{\nu_2-\nu_1}{\nu_4-\nu_2}\sn^2(\theta,m)},
\end{equation}
where it is assumed that $\al I(0)=\nu_1$,
\begin{equation}\label{eq21}
    \theta=\sqrt{(\nu_3-\nu_1)(\nu_4-\nu_2)}\,\xi/2,
\end{equation}
\begin{equation}\label{eq22}
    m=\frac{(\nu_4-\nu_3)(I_\nu-\nu_1)}{(\nu_4-\nu_2)(\nu_3-\nu_1)},
\end{equation}
$\cn$ and $\sn$ being Jacobi elliptic functions \cite{AbramowitzStegun-72}.
The period of the oscillating with change of $t$ function \eqref{eq20} is
\begin{equation}\label{eq23}
    T=\frac{4K(m)}{\sqrt{(\nu_3-\nu_1)(\nu_4-\nu_2)}}=\frac{K(m)}{\sqrt{(\la_3^2-\la_1^2)(\la_4^2-\la_2^2)}},
\end{equation}
where $K(m)$ is the complete elliptic integral of the first kind
\cite{AbramowitzStegun-72}.

In the limit $\nu_3\to \nu_2$ ($m\to1$) the period
tends to infinity and the solution (\ref{eq20}) acquires the soliton form
\begin{equation}\label{eq24}
    \al I=\nu_2-\frac{\nu_2-\nu_1}{\cosh^2\theta+\frac{\nu_2-\nu_1}{\nu_4-\nu_2}\sinh^2\theta}.
\end{equation}
This is a ``dark soliton'' for the variable $I$.

The limit $m\to0$ can be reached in two ways.

(i) If $\nu_2\to \nu_1$, then the solution transforms into a linear harmonic
wave
\begin{equation}\label{eq25}
    \begin{split}
    \al I&\cong \nu_2-\frac12(\nu_2-\nu_1)\cos(\om\xi),\\
    \om&=\sqrt{(\nu_3-\nu_1)(\nu_4-\nu_1)}.
\end{split}
\end{equation}

(ii) If $\nu_4=\nu_3$ but $\nu_1\neq \nu_2$, then we arrive at
the nonlinear trigonometric solution:
\begin{equation}\label{eq26}
    \begin{split}
    \al I&=\nu_2-\frac{(\nu_2-\nu_1)\cos^2\theta}{1+\frac{\nu_2-\nu_1}{\nu_3-\nu_2}\sin^2\theta},\\
    \theta&=\sqrt{(\nu_3-\nu_1)(\nu_3-\nu_2)}\,\xi/2.
    \end{split}
\end{equation}
If we take the limit $\nu_2-\nu_1\ll \nu_3-\nu_1$ in this solution, then we
return to the small-amplitude limit (\ref{eq25}) with $\nu_4=\nu_3$. On
the other hand, if we take here the limit $\nu_2\to \nu_3=\nu_4$, then the
argument of the trigonometric functions becomes small and we can
approximate them by the first terms of their series expansions. This
corresponds to an algebraic soliton of the form
\begin{equation}\label{eq27}
   \al I=\nu_2-\frac{\nu_2-\nu_1}{1+(\nu_2-\nu_1)^2\xi^2/4}.
\end{equation}

(B) In the second case, the variable $\al I$ oscillates in the interval
\begin{equation}\label{eq28}
    \nu_3\leq \al I\leq \nu_4\; .
\end{equation}
Here again, a standard calculation yields
\begin{equation}\label{eq30}
   \al I=\nu_3+\frac{(\nu_4-\nu_3)\cn^2(\theta,m)}{1+\frac{\nu_4-\nu_3}{\nu_3-\nu_1}\sn^2(\theta,m)}
\end{equation}
with the same definitions (\ref{eq21}), (\ref{eq22}), and (\ref{eq23})
for $\theta$, $m$, and $T$, correspondingly. In this case we have $\al I(0)=\nu_4$.
In the soliton limit $\nu_3\to \nu_2$ ($m\to1$) we get
\begin{equation}\label{eq31}
   \al I=\nu_2+\frac{\nu_4-\nu_2}{\cosh^2\theta+\frac{\nu_4-\nu_2}{\nu_2-\nu_1}\sinh^2\theta}.
\end{equation}
This is a ``bright soliton'' for the variable $I$.

Again, the limit $m\to0$ can be reached in two ways.

(i) If $\nu_4\to \nu_3$, then we obtain a small-amplitude harmonic wave
\begin{equation}\label{eq32}
    \begin{split}
   \al I&\cong \nu_3+\frac12(\nu_4-\nu_3)\cos(\om\xi),\\
    \om&=\sqrt{(\nu_3-\nu_1)(\nu_3-\nu_2)}.
    \end{split}
\end{equation}

(ii) If $\nu_2=\nu_1$, then we obtain another nonlinear trigonometric solution,
\begin{equation}\label{eq33}
    \begin{split}
    \al I&=\nu_3+\frac{(\nu_4-\nu_3)\cos^2\theta}{1+\frac{\nu_4-\nu_3}{\nu_3-\nu_1}\sin^2\theta},\\
    \theta&=\sqrt{(\nu_3-\nu_1)(\nu_4-\nu_1)}\,\xi/2.
    \end{split}
\end{equation}
If we assume that $\nu_4-\nu_3\ll \nu_4-\nu_1$, then this reproduce
the small-amplitude limit (\ref{eq32}) with $\nu_2=\nu_1$. On the other hand,
in the limit $\nu_3\to \nu_2=\nu_1$ we obtain the algebraic soliton solution:
\begin{equation}\label{eq34}
    \al I=\nu_1+\frac{\nu_4-\nu_1}{1+(\nu_4-\nu_1)^2\xi^2/4}.
\end{equation}

The convenience of this form of periodic solutions of our equation is related with the fact that
the parameters $\lambda_i$, connected with $\nu_i$ by the formulae (\ref{zeros1}), (\ref{zeros2}),
play the role of Riemann invariants in Whitham theory of modulations. For both cases (\ref{zeros1}), (\ref{zeros2})
we have the identities
\begin{equation}\label{}
    m=\frac{(\nu_4-\nu_3)(\nu_2-\nu_1)}{(\nu_4-\nu_2)(\nu_3-\nu_1)}
        =\frac{(\lambda_4^2-\lambda_3^2)(\lambda_2^2-\lambda_1^2)}{(\lambda_4^2-\lambda_2^2)(\lambda_3^2-\lambda_1^2)}.
\end{equation}

Now we shall consider slowly modulated waves. In
this case, the parameters $\lambda_i$ ($i = 1, 2, 3, 4$) become slowly
varying functions of $x$ and $t$ changing little in one period and, as was found
in Ref.~\cite{kamch-1990}, they can serve as Riemann invariants.
Evolution of $\la_i$ is governed by the Whitham modulation equations
\begin{equation}\label{WhithamEq}
    \frac{\partial \lambda_i}{\partial x}+\frac{1}{v_i}\frac{\partial \lambda_i}{\partial t}=0.
\end{equation}
The inverse Whitham velocities appearing in these equations can
be computed by means of the formulae
\begin{equation}\label{}
    \begin{split}
    \frac{1}{v_i}=\left(1-\frac{T}{\partial_i T}\partial_i\right)\frac{1}{V},
    \quad \mbox{where} \quad \partial_i\equiv\frac{\partial}{\partial \lambda^2_i},
    \end{split}
\end{equation}
with the use of equations (\ref{phaseV}), (\ref{eq23}). Hence, a simple calculation yields
the explicit expressions
\begin{equation}\label{}
    \begin{split}
    & \frac{1}{v_1}=\frac{1}{\al}-\frac{1}{2}\sum_{i=1}^{4}\lambda_i^2
        -\frac{(\lambda_4^2-\lambda_1^2)(\lambda_2^2-\lambda_1^2)K(m)}
        {(\lambda_4^2-\lambda_1^2)K(m)-(\lambda_4^2-\lambda_2^2)E(m)}, \\
    & \frac{1}{v_2}=\frac{1}{\al}-\frac{1}{2}\sum_{i=1}^{4}\lambda_i^2
        +\frac{(\lambda_3^2-\lambda_2^2)(\lambda_2^2-\lambda_1^2)K(m)}
        {(\lambda_3^2-\lambda_2^2)K(m)-(\lambda_3^2-\lambda_1^2)E(m)}, \\
    & \frac{1}{v_3}=\frac{1}{\al}-\frac{1}{2}\sum_{i=1}^{4}\lambda_i^2
        -\frac{(\lambda_4^2-\lambda_3^2)(\lambda_3^2-\lambda_2^2)K(m)}
        {(\lambda_3^2-\lambda_2^2)K(m)-(\lambda_4^2-\lambda_2^2)E(m)}, \\
    & \frac{1}{v_4}=\frac{1}{\al}-\frac{1}{2}\sum_{i=1}^{4}\lambda_i^2
        +\frac{(\lambda_4^2-\lambda_2^2)(\lambda_4^2-\lambda_1^2)K(m)}
        {(\lambda_4^2-\lambda_1^2)K(m)-(\lambda_3^2-\lambda_1^2)E(m)}.
    \end{split}
\end{equation}

In a modulated wave representing a dispersive shock wave, the Riemann invariants change
with $t$ and $x$. The dispersive shock wave occupies a time interval at whose edges
two of the Riemann invariants coincide.
The soliton edge corresponds to $\la_3=\la_2$ $(m=1)$ and at this edge
the Whitham velocities are given by
\begin{equation}\label{sol-limit}
     \begin{split}
    \frac{1}{v_1} & =\frac{1}{\al}-(3\lambda_1^2+\lambda_4^2), \quad
    \frac{1}{v_4}  =\frac{1}{\al}-(\lambda_1^2+3\lambda_4^2),\\
    \frac{1}{v_2} & =\frac{1}{v_3}=\frac{1}{\al}-(\lambda_1^2+2\lambda_2^2+\lambda_4^2).
    \end{split}
\end{equation}
The small amplitude limit $m=0$ can be obtained in two ways. If
$\la_3=\la_4$, then we get
\begin{equation}\label{small-limit}
\begin{split}
    \frac{1}{v_1} & =\frac{1}{\al}-(3\lambda_1^2+\lambda_2^2), \quad \frac{1}{v_2}=\frac{1}{\al}-(\lambda_1^2+3\lambda_2^2), \\
    \frac{1}{v_3} & =\frac{1}{v_4}=\frac{1}{\al}-4\lambda_4^2-\frac{(\lambda_2^2-\lambda_1^2)^2}{\lambda_1^2-\lambda_2^2-2\lambda_4^2},
    \end{split}
\end{equation}
and if $\la_2=\la_1$, then
\begin{equation}\label{small-limit2}
\begin{split}
    \frac{1}{v_1} & =\frac{1}{v_2}=\frac{1}{\al}-4\lambda_1^2-\frac{(\lambda_4^2-\lambda_3^2)^2}{\lambda_3^2+\lambda_4^2-2\lambda_1^2}, \\
    \frac{1}{v_3} & =\frac{1}{\al}-(3\lambda_3^2+\lambda_4^2), \quad \frac{1}{v_4}=\frac{1}{\al}-(\lambda_3^2+3\lambda_4^2).
    \end{split}
\end{equation}

Now we are ready to discuss the key elements from which  any wave
structure evolving from an initial discontinuity consists.

\section{Elementary wave structures}\label{sec5}

Let the initial (input) conditions have a step-like form,
\begin{equation}\label{init}
    \begin{split}
    I(x=0)=
    \begin{cases}
        I^L, & \quad t<0, \\
        I^R, & \quad t>0,
    \end{cases}\quad \\
        u(x=0)=
    \begin{cases}
        u^L, & \quad t<0, \\
        u^R, & \quad t>0.
    \end{cases}\quad
    \end{split}
\end{equation}
Evolution of this step-like pulse leads to formation of quite complex structures
consisting of simpler elements. We shall describe these elements in the
present section.

\subsection{Rarefaction waves}

For smooth enough wave patterns we can neglect the last dispersion term
in the second equation of the system (\ref{Hydro}) and arrive at
the so-called dispersionless equations (\ref{Hydro2}).
First of all, this system admits a trivial solution for which
$I=\mathrm{const}$ and $u=\mathrm{const}$. We shall call such a solution
a ``plateau''. Introducing the Riemann invariants (\ref{RiemannInv}),
we transform the hydrodynamic equations (\ref{Hydro2})
to the diagonal form
\begin{equation}\label{RiemannEquat}
    \begin{split}
    \frac{\partial r_{\pm}}{\partial x}+\frac{1}{v_{\pm}}\frac{\partial r_{\pm}}{\partial t}=0,
    \end{split}
\end{equation}
where the Riemann velocities are expressed via the Riemann invariants by the relations
\begin{equation}\label{riemann-vel}
    \begin{split}
    \frac{1}{v_+}=\frac32r_{+}+\frac12r_{-},\quad  \frac{1}{v_-}=\frac12r_{+}+\frac32r_{-}.
    \end{split}
\end{equation}
In terms of $I$ and $u$ these velocities are given by Eqs.~(\ref{211.4b}).
It is clear that the system is modulationally unstable if
\begin{equation}\label{ModulUnstable}
    \begin{split}
    u>\frac{1}{\al}+\al I.
    \end{split}
\end{equation}

\begin{figure}[t] \centering
\includegraphics[width=6cm]{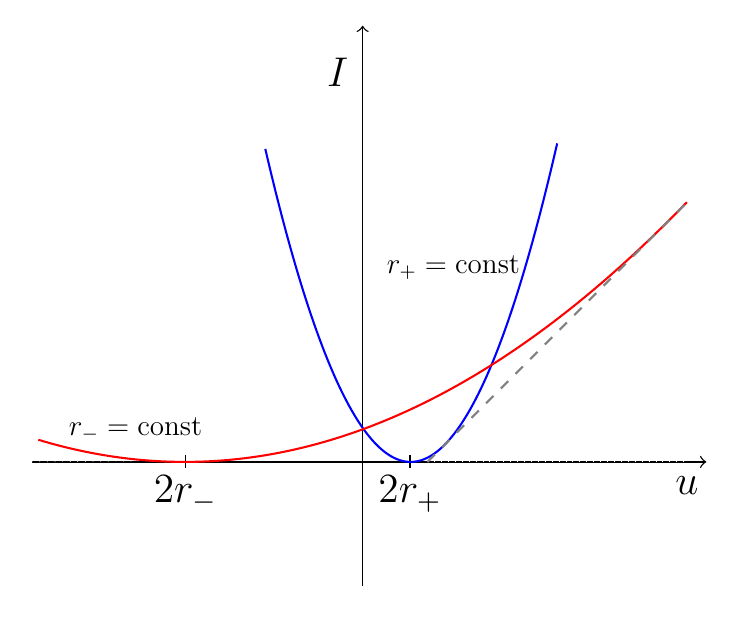}
\caption{Relation between $u$ and $I$ for simple wave solutions in the dispersionless regime.
One line corresponds to $r_-=\mathrm{const}$, and another one to $r_+=\mathrm{const}$.
Dashed gray area shows the modulationally unstable region.}
\label{Fig2}
\end{figure}

A rarefaction wave belongs to the class of simple wave solutions discussed in section \ref{sec3}
and it is characterized
by the condition that one of the Riemann invariants has a constant value along the flow,
$r_+=\mathrm{const}$ or $r_-=\mathrm{const}$.
Consequently, according to the definition (\ref{RiemannInv}), these simple wave solutions are
represented in the $(u,I)$-plane by the parabolas
\begin{equation}\label{}
    \begin{split}
    I=\frac{(u-2r_\pm)^2}{4(1-2\al r_\pm)}.
    \end{split}
\end{equation}
To have the
intensity positive, it is necessary to fulfil the condition $r_-\leq r_+\leq 1/(2\al)$.
By virtue of obvious inequality $r_+\geq r_-$ the parabola corresponding to
$r_-=\mathrm{const}$ has greater curvature than the parabola for $r_+=\mathrm{const}$ (see Fig.~\ref{Fig2}).
Both parabolas touch the boundary line $I=u/\al-1/\al^2$ of the instability region (in fact,
this line is an envelope of a pencil of parabolas $I=(u-2r)^2/(4(1-2\al r))$ with $r$ as a parameter).
In the Fig.~\ref{Fig2}, the modulationally unstable region (\ref{ModulUnstable}) is dashed.
Along the line $u=1/\al$ both derivatives $\prt r_+/\prt u=0$, $\prt r_+/\prt I=0$ vanish
and $r_+$ reaches here its maximal value equal to $r_+=1/(2\al)$. We say that the line $u=1/\al$
separates two monotonicity regions $u<1/\al$ and $u>1/\alpha$ in the half-plane $I\geq0$.
The two intersection points of parabolas correspond to uniform flows with
constant parameters $I=\mathrm{const}$ and $u=\mathrm{const}$, that is to
the plateau solutions.
It is easy to express the physical variables $I$ and $u$ in terms of $r_-$, $r_+$,
\begin{equation}\label{IandUbyR}
    \begin{split}
    I & =\frac{1}{2\al^2}\left(1-\al(r_++r_-)\pm\sqrt{(1-2\al r_+)(1-2\al r_-)}\right), \\
    u & =\frac{1}{\al}\left(1\pm\sqrt{(1-2\al r_+)(1-2\al r_-)}\right).
    \end{split}
\end{equation}

The initial conditions (\ref{init}) do not contain any parameters with dimension of time or length.
Therefore solutions of equations (\ref{RiemannEquat}) can depend on the self-similar variable
$\tau=t/x$ only, that is $r_{\pm}=r_{\pm}(\tau)$, and then this system reduces to
\begin{equation}\label{}
    \begin{split}
    \left(\frac{1}{v_-}-\tau\right)\frac{dr_-}{d\tau}=0, \quad \left(\frac{1}{v_+}-\tau\right)\frac{dr_+}{d\tau}=0.
    \end{split}
\end{equation}
We note again that these equations have a simple
solution $r_-=\mathrm{const}$,
$r_+=\mathrm{const}$ with constant $u$ and $I$ which corresponds
to the mentioned above plateau region.

\begin{figure}[t] \centering
\includegraphics[width=6cm]{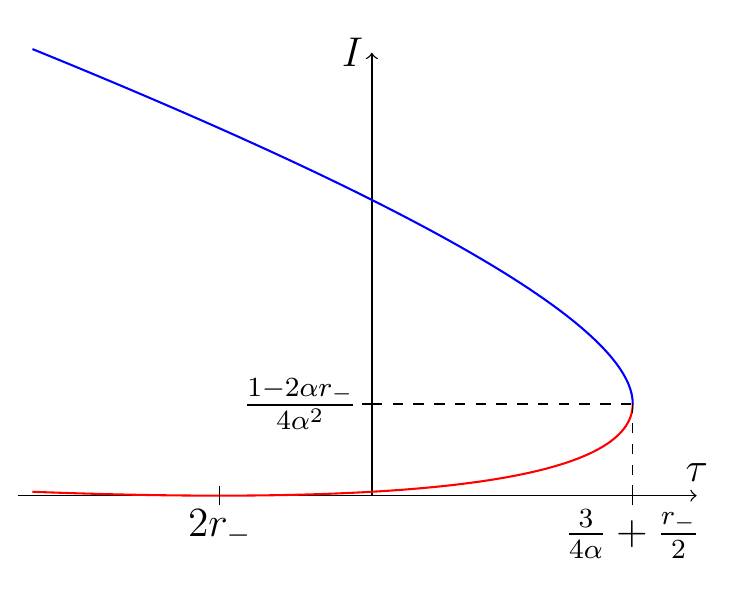}
\caption{Dependence of simple-wave solution $I(\tau)$  on $\tau=t/x$.
The upper sign in Eq.~(\ref{I_tau}) corresponds to the upper branch curve and the lower one to the lower branch.}
\label{Fig3}
\end{figure}

\begin{figure}[t] \centering
\includegraphics[width=6cm]{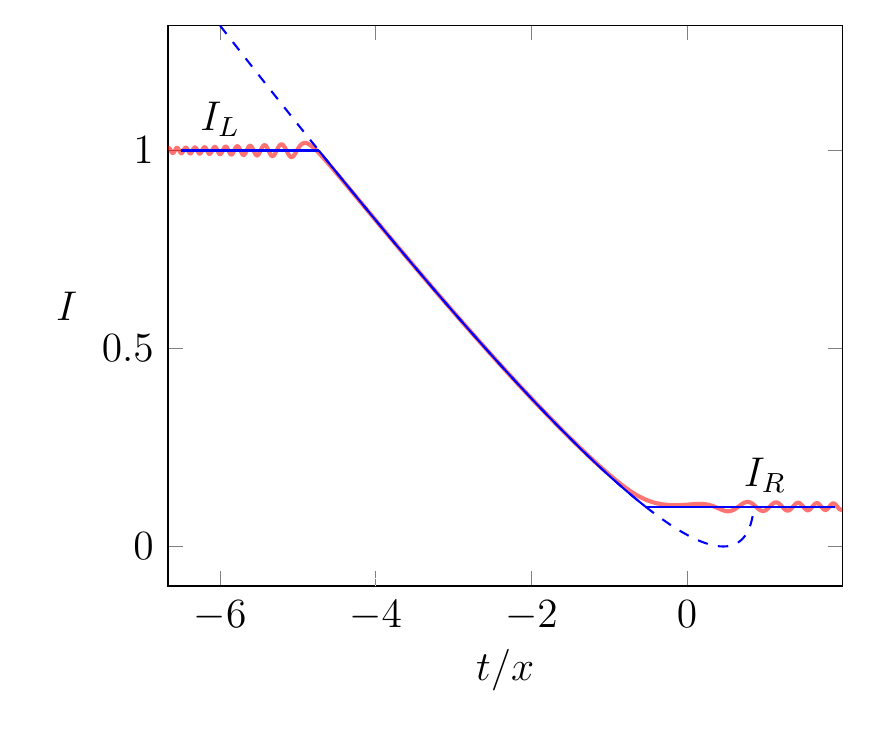}
\caption{Example of the simple-wave solution for $I^L=1$, $u^L=-1$, $I^R=0.1$, $u^R=0$, $r_+=0.232$.
Numerical solution of the mNLS equation (\ref{DNLS1}) is shown in red, analytical approximation is shown
in blue.}
\label{Fig4}
\end{figure}

Turning to self-similar simple wave solutions, let us consider for definiteness
the case when $r_{-}=\mathrm{const}$. Then we have
\begin{equation}\label{}
    \begin{split}
    \frac{1}{v_+} & =u-2\al I+\sqrt{I\left(\al^2I-\al u+1\right)}=\tau=\frac{t}{x}, \\
    r_{-} & =\frac{u}{2}-\al I-\sqrt{I\left(\al^2I-\al u+1\right)}=\mathrm{const}.
    \end{split}
\end{equation}
Solving this system with respect to $I$ and $u$ yields
\begin{equation}\label{I_tau}
    \begin{split}
    I(\tau)&=\frac{1}{2\al^2}-\frac{1}{3\al}(r_-+\tau)\\
    &\pm\frac{1}{2\al^2} \sqrt{\left(1-2\al r_-\right)\left[1+\frac{2}{3}\al(r_--2\tau)\right]},\\
     u(\tau)&=\frac23\left[\tau+r_-+3\alpha I(\tau)\right].
    \end{split}
\end{equation}
Plots for the intensity for both choices of the sign are shown in the Fig.~\ref{Fig3}.
We see that in the self-similar solutions the variable $\tau$ must be below its maximum value
\begin{equation}\label{}
    \begin{split}
    \tau\leq\frac{3}{4\al}+\frac{r_-}{2},
    \end{split}
\end{equation}
at which the solutions coincide with each other and the intensity assumes
the common value equal to
\begin{equation}\label{}
    \begin{split}
    I=\frac{1-2\al r_-}{4\al^2}.
    \end{split}
\end{equation}
This means that both types of solutions (lower or upper branchs) must match some other
element of the whole structure at $\tau$ smaller than its maximal possible value.
Similar formulas and plots can be obtained for the solution
$r_+=\mathrm{const},v_-(r_-,r_+)=x/t\equiv 1/\tau$.
This wave configuration represents a rarefaction wave.
In the general case this type of wave can connect uniform
flows with equal values of the corresponding Riemann invariants
$r_-^L=r_-^R$ or $r_+^L=r_+^R$. Example of corresponding
distribution is shown in Fig.~\ref{Fig4}. The analytical simple
wave approximation agrees with the exact numerical solution
very well.

\begin{figure}[t] \centering
\includegraphics[width=8cm]{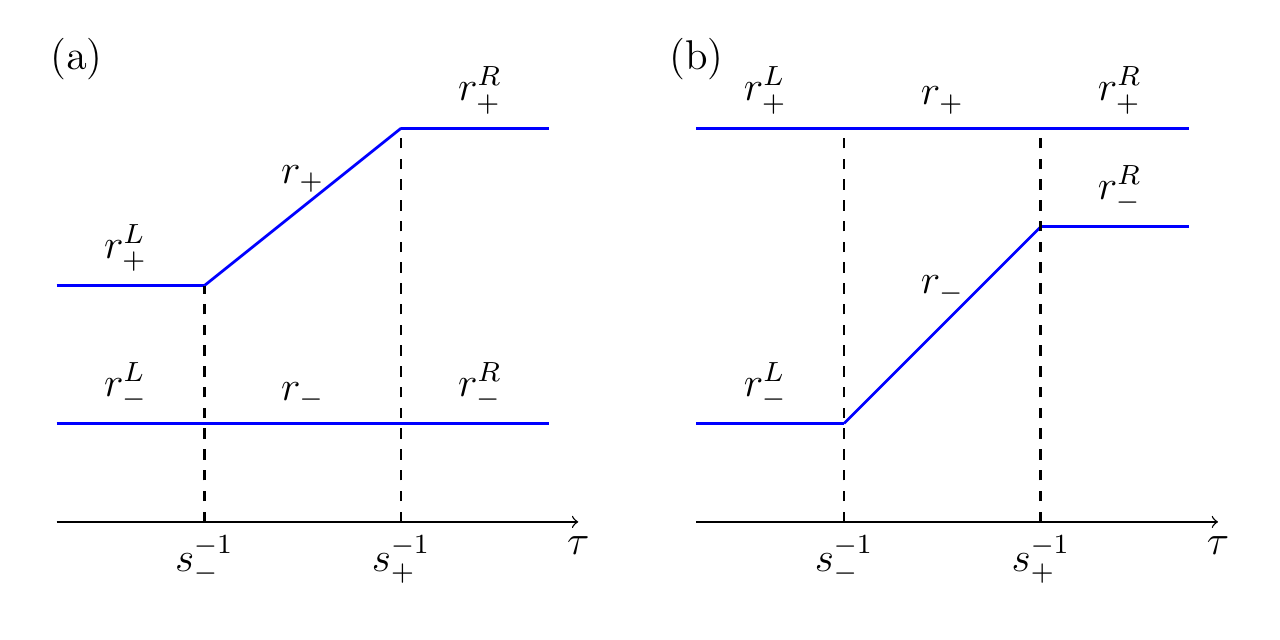}
\caption{Diagrams representing the evolution of the Riemann invariants as functions
of $\tau=t/x$ in the rarefaction wave solutions of the hydrodynamic equations:
(a) $r_-=\mathrm{const}$, $r_+^L<r_+^R$; (b) $r_+=\mathrm{const}$, $r_-^L<r_-^R$.}
\label{Fig5}
\end{figure}

Both branches in Fig.~\ref{Fig3} correspond to the same solution
of the equations (\ref{RiemannEquat}) written for the Riemann invariants,
\begin{equation}\label{300.7}
  r_-=r_-^0=\mathrm{const},\quad \frac1{v_+}=\frac32r_-^0+\frac12r_+=\tau=\frac{t}x,
\end{equation}
and two solutions appear due to the two-valued character of the formulae (\ref{IandUbyR})
and (\ref{I_tau}).
Thus, in these self-similar solutions one of the Riemann invariants
must be constant and another one must increase with $\tau$
according to Eqs.~(\ref{300.7}). The dependence of the Riemann
invariants on the physical parameters must also be
monotonous in order to keep the solution single-valued.
Hence both edge points of the rarefaction wave must
lie either on the left or on the right side of the line $u=1/\al$,
along which the Riemann invariants reach their
extremal values.  As was mentioned above, we shall call the two regions
on the left and right sides of this line as monotonicity regions.
Dependence of the Riemann invariants on $\tau$ is sketched in Fig.~\ref{Fig5}
for two possible situations with $r_-$ or $r_+$ constant.
The edge velocities of these rarefaction waves are equal to
\begin{equation}\label{}
    \begin{split}
    & (a) \quad s_-^{-1}=\frac{1}{2}r_-^L+\frac{3}{2}r_+^L,
        \quad s_+^{-1}=\frac{1}{2}r_-^R+\frac{3}{2}r_+^R;\\
    & (b) \quad s_-^{-1}=\frac{3}{2}r_-^L+\frac{1}{2}r_+^L,
        \quad s_+^{-1}=\frac{3}{2}r_-^R+\frac{1}{2}r_+^R.
    \end{split}
\end{equation}
Obviously, the corresponding wave structures must satisfy the conditions (a) $r_+^L<r_+^R$,
$r_-^L=r_-^R$ or (b) $r_+^L = r_+^R$, $r_-^L<r_-^R$. It is natural to ask,
what happens if we have the initial conditions satisfying opposite inequalities,
and to answer this question we have to consider the DSW structures.

\subsection{Cnoidal dispersive shock waves}

\begin{figure}[t] \centering
\includegraphics[width=8cm]{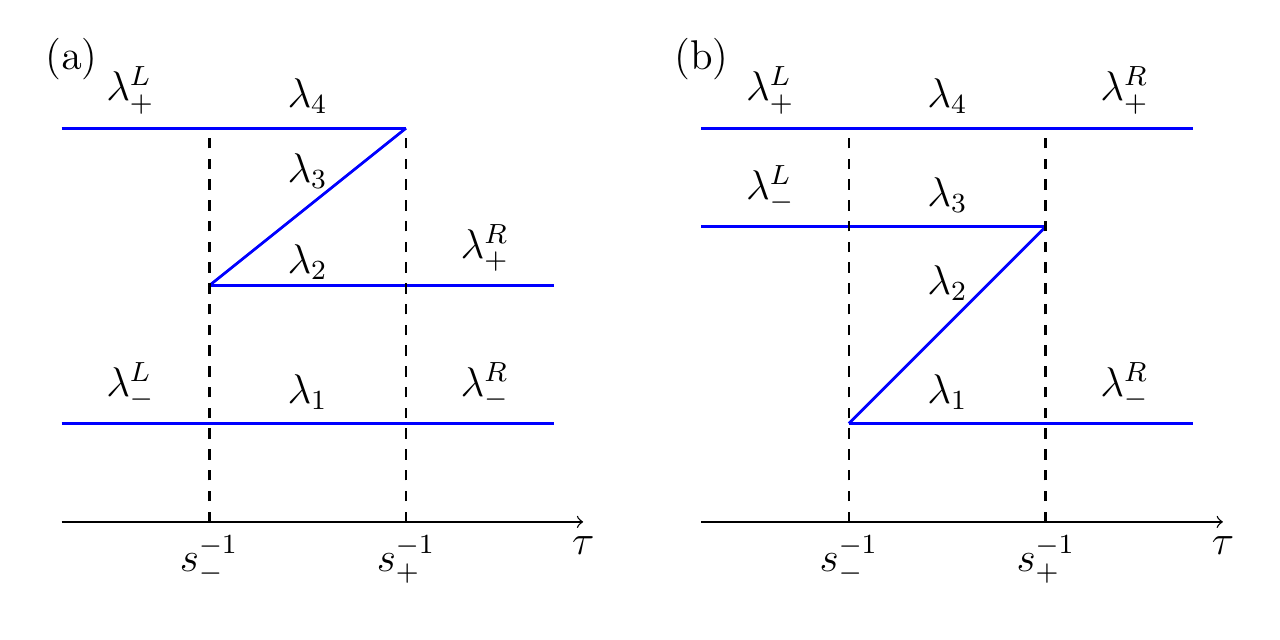}
\caption{Diagrams representing the dependence of the Riemann invariants on $\tau=t/x$
in dispersive shock wave solutions of the Whitham equations: (a) $\la_-^L=\la_-^R$, $\la_+^L>\la_+^R$;
(b) $\la_+^L=\la_+^R$, $\la_-^L>\la_-^R$.}
\label{Fig6}
\end{figure}

The other two possible solutions of Eqs.~(\ref{RiemannEquat}) are sketched
in Fig.~\ref{Fig6}, where for future convenience we have made
the change $r\mapsto\la$ ($r_\pm$ will be functions of $\la_\pm$ defined below),
and they satisfy the boundary conditions (a) $\la_+^L = \la_+^R$,
$\la_-^L > \la_-^R$ or (b) $\la_+^L > \la_+^R$, $\la_-^L = \la_-^R$. In the
dispersionless approximation these multi-valued solutions are nonphysical.
However, we can give them clear physical sense by understanding $\la_i$ as four
Riemann invariants of the Whitham system that describe evolution of a
modulated nonlinear periodic wave. We interpret this as formation of
cnoidal dispersive shock wave from the initial discontinuity with such a type
of the boundary conditions.

To find the solution of
equations (\ref{WhithamEq}), we use again the argument that the wavelength of
the DSW is negligibly small compared with the large scale of the whole structure
and at this scale the initial conditions for the Whitham equations (\ref{WhithamEq}),
which are similar to the Riemann equations (\ref{RiemannEquat}),
do not contain parameters with dimension of length, so the modulation
parameters depend on the self-similar variable $\tau=t/x$ only.
Therefore, equations (\ref{WhithamEq}) reduce to
\begin{equation}\label{DispSelf-simEq}
    \left(\frac{1}{v_i}-\tau\right)\frac{d\la_i}{d\tau}=0.
\end{equation}
Hence we find again that only one Riemann invariant varies along the DSW, while
the other three are constant, that is the corresponding diagram reproduces the picture shown in
Fig.~\ref{Fig6}. The limiting expressions (\ref{sol-limit})  for
the Whitham velocities must coincide with expressions (\ref{riemann-vel}) for dispersionless
Riemann velocities and therefore we can relate the corresponding dispersionless and dispersive
Riemann invariants by the formulae
\begin{equation}\label{}
    \begin{split}
    (a) & \quad \lambda_-^L =-\sqrt{\frac{1}{4\al}-\frac{r_-^L}{2}}, \quad
         \lambda_+^L=-\sqrt{\frac{1}{4\al}-\frac{r_+^L}{2}}, \\
    (b) & \quad \lambda_-^R  =-\sqrt{\frac{1}{4\al}-\frac{r_-^R}{2}}, \quad
         \lambda_+^R=-\sqrt{\frac{1}{4\al}-\frac{r_+^R}{2}}
    \end{split}
\end{equation}
at the soliton edges of the DSW.
Here ${r}_{\pm}^{L,R}$ are the Riemann invariants of the dispersionless
theory that are defined by Eqs.~(\ref{RiemannInv}). They describe
the plateau solution at the soliton edge of the DSW. In a similar way, at the small-amplitude
edges we find similar relations
\begin{equation}\label{}
    \begin{split}
    (a) \quad \lambda_-^R & =-\sqrt{\frac{1}{4\alpha}-\frac{r_-^R}{2}}, \quad
    \lambda_+^R=-\sqrt{\frac{1}{4\alpha}-\frac{r_+^R}{2}},
    \end{split}
\end{equation}
and
\begin{equation}\label{}
   \begin{split}
    (b) \quad \lambda_-^L & =-\sqrt{\frac{1}{4\al}-\frac{r_-^L}{2}}, \quad
    \lambda_+^L=-\sqrt{\frac{1}{4\al}-\frac{r_+^L}{2}}.
    \end{split}
\end{equation}
Again the limiting expressions (\ref{small-limit}) and (\ref{small-limit2})
coincide with the dispersionless expressions (\ref{riemann-vel}).
Then the self-similar solutions of the Whitham equations (\ref{DispSelf-simEq})
are given by
\begin{equation}\label{RInv-sols}
    \begin{split}
    &(a) \quad v_3^{-1}(\la_-^L,\la_+^R ,\la_3(\tau),\la_+^L)=\tau\; ;\\
    &\mbox{or}\\
    &(b) \quad v_2^{-1}(\la_-^R,\la_2(\tau),\la_-^L,\la_+^L)=\tau\; ,
  \end{split}
\end{equation}
which define the dependence of the Riemann invariants (modulation parameters) $\la_3$ or $\la_2$
on $\tau$ in implicit form.
The edges of the DSW propagate with velocities
\begin{equation}\label{}
    \begin{split}
    (a) & \quad s_-^{-1}=\frac{1}{\al}-\left((\lambda_-^{L})^2+2(\lambda_+^{R})^2+(\lambda_+^{L})^2\right)\; , \\
        & \quad s_+^{-1}=\frac{1}{\al}-4(\lambda_+^{L})^2-
            \frac{((\lambda_+^R)^2-(\lambda_-^R)^2)^2}{(\lambda_+^R)^2+(\lambda_-^R)^2-2(\lambda_+^L)^2}\; ;\\
    (b) & \quad s_-^{-1}=\frac{1}{\al}-4(\lambda_+^R)^2-
        \frac{((\lambda_+^L)^2-(\lambda_-^L)^2)^2}{(\lambda_+^L)^2+(\lambda_-^L)^2-2(\lambda_+^R)^2}\; ,\\
        & \quad s_+^{-1}=\frac{1}{\al}-\left((\lambda_-^R)^2+2(\lambda_+^L)^2+(\lambda_+^R)^2\right)\; .
    \end{split}
\end{equation}

\begin{figure}[t] \centering
\includegraphics[width=7cm]{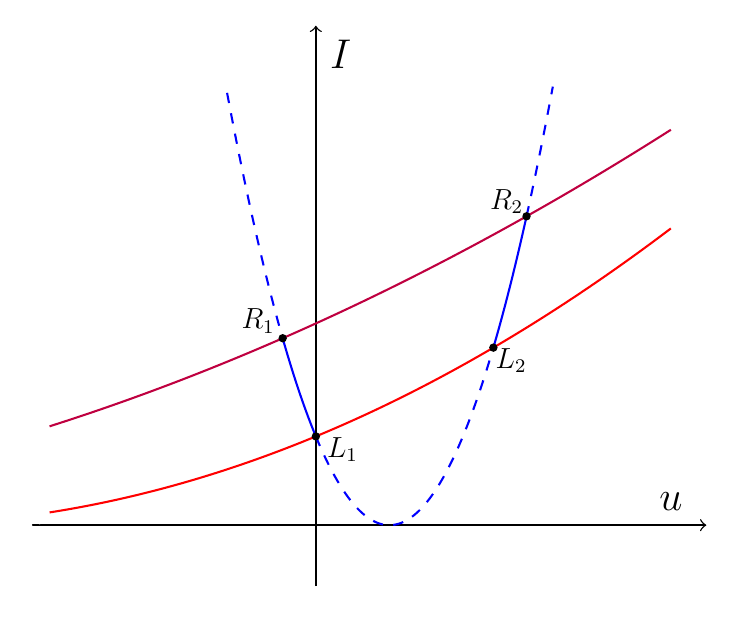}
\caption{Example of two possible paths in the ($u,I$)-plane between the left and right
boundary for the case of dispersive shock waves. To satisfy the given boundary conditions, the certain
path must be chosen: for the path $L_1 \rightarrow R_1$ formulae (\ref{zeros1})
and for the path $L_2 \rightarrow R_2$ formulae (\ref{zeros2}) are used.
Corresponding wave structures are shown in Fig.~\ref{Fig8} and they satisfy the same solution
of the Whitham equations, but different boundary conditions in physical variables.
               }
\label{Fig7}
\end{figure}

\begin{figure}[t] \centering
    \includegraphics[width=4cm]{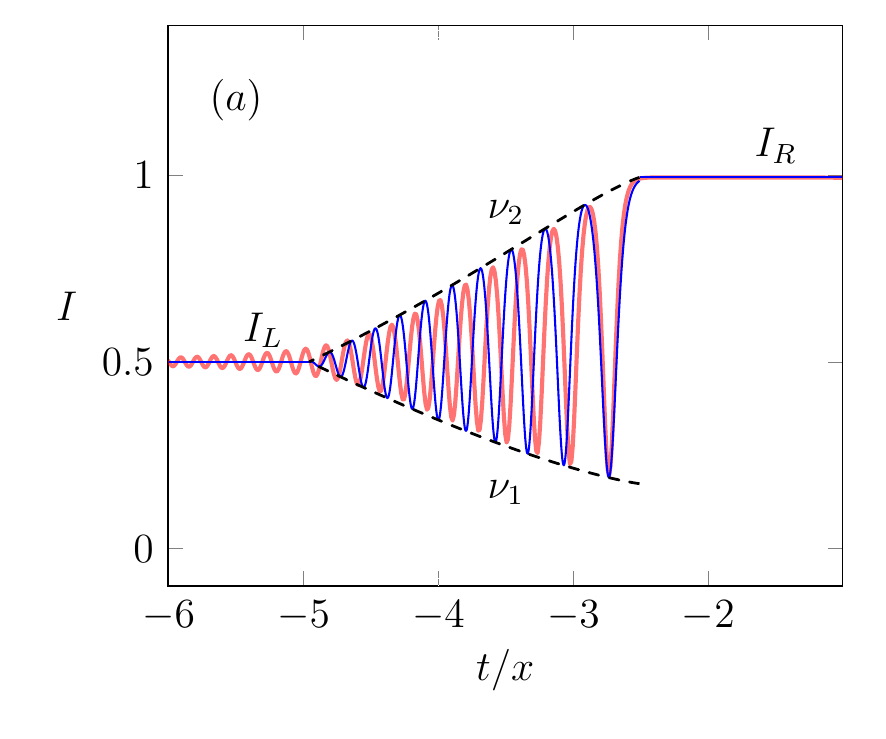}
    \includegraphics[width=4cm]{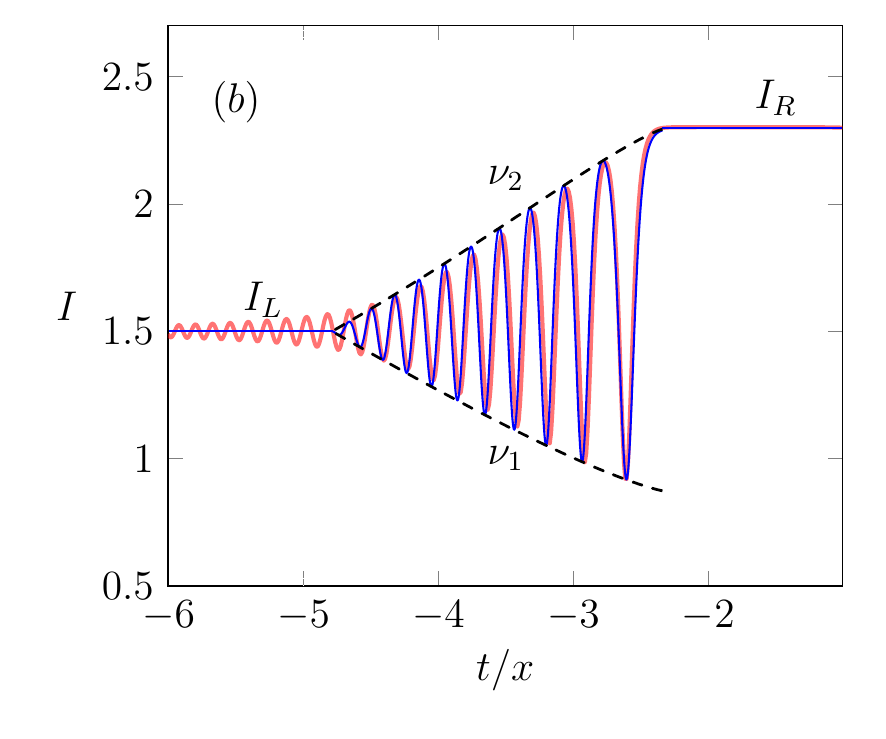}
    \caption{Comparison of analytical and numerical solutions of the mNLS equation (\ref{DNLS1})
    for two different boundary conditions and the same solution of the Whitham equations for
    the modulation parameters: (a) $I^L=0.5$, $u^L=0$, $I^R=0.99$, $u^R=-0.3$;
    (b) $I^L=1.5$, $u^L=2$, $I^R=2.29$, $u^R=2.3$ with $\alpha=1$. The Riemann invariants
    are equal to $r_+^L=r_+^R=0.37$, $r_-^L=-1.37$ and $r_-^R=-2.65$. Thin line corresponds to the analytic
    solution, thick gray line to numerics, dashed lines show analytical envelopes.}
    \label{Fig8}
\end{figure}

It should be stressed that each $\la$-diagram in Fig.~\ref{Fig6} corresponds to
two different dispersive shock waves, because we have two
mappings (\ref{zeros1}) and (\ref{zeros2}) from Riemann invariants to the physical
parameters. This point will be important in classification of
the wave structures evolving from the initial discontinuities.
For example, let us consider the case (b) ($\la_+^L=\la_+^R$, $\la_-^L>\la_-^R$)
(the diagram Fig.~\ref{Fig6}(b)).
In Fig.~\ref{Fig7} the parabolas of constant Riemann invariants in the $(u,I)$-plane are shown.
We see that there are two paths
$L_1 \rightarrow R_1$ and $L_2 \rightarrow R_2$ which connect pairs of points
with the same values of both Riemann invariants.
The points $L_1$ and $L_2$ correspond to the left boundary
condition with the Riemann invariants equal to $\la_-^L$ and $\la_+^L$,
and the points $R_1$ and $R_2$ correspond to the right
boundary condition with the Riemann invariants equal to $\la_-^R$ and $\la_+^R$ ($\la_+^R=\la_+^L$).
These paths are described correspondingly by the maps (\ref{zeros1}) or (\ref{zeros2})
of Riemann invariants to the parameters $\nu$ that parameterize the periodic solutions.
Substitution of the solutions (\ref{RInv-sols}) for the Riemann invariants in the
formulae (\ref{zeros1}) and (\ref{zeros2}) with the use of (\ref{eq20}) yields
the $\tau$-dependence of the parameters in the modulated periodic solutions resulting in the DSW structure.
In Fig.~\ref{Fig8} we compare the numerical and analytical approximate solution
for the DSW with the constant Riemann invariant $\la_+$.
For the path $L_1 \rightarrow R_1$ zeros (\ref{zeros1}) are used,
and for the path $L_2 \rightarrow R_2$ we use formulas (\ref{zeros2}).

In a similar way, the diagram Fig.~\ref{Fig6}(a)
produces two other wave structures.

\subsection{Contact dispersive shock wave}

\begin{figure}[t] \centering
\includegraphics[width=7cm]{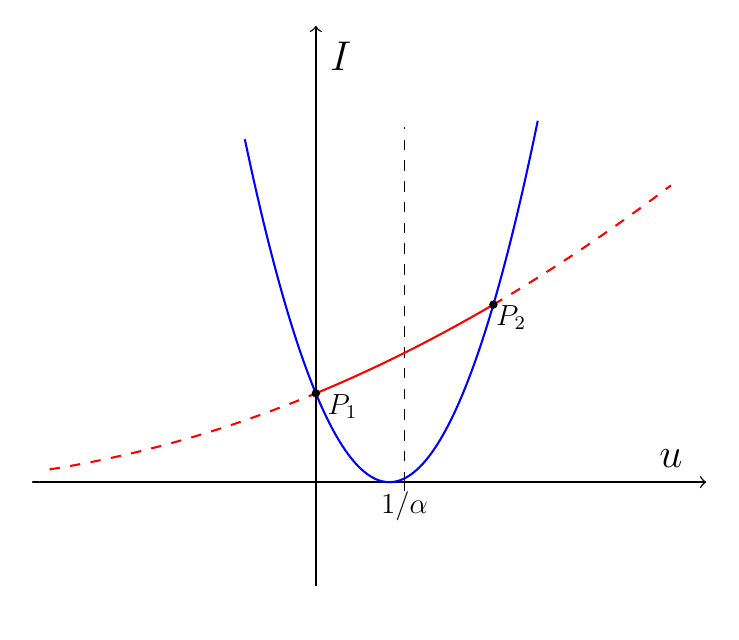}
\caption{Example of path for a contact DSW ($r_+^L=r_+^R$ and $r_-^L=r_-^R$).
The $r_-$-parabola crosses the vertical $u=1/\al$ separating the regions of monotonicity.
Two directions $P_1 \leftrightarrow P_2$ are described by the mappings (\ref{zeros1}) and
(\ref{zeros2}). Corresponding $\la$-diagrams are shown in Fig.~\ref{Fig10} and the wave structures
are shown in Fig.~\ref{Fig11}.}
\label{Fig9}
\end{figure}

We now consider the situation in which the Riemann
invariants have equal values at both edges of the shock, i.e., when
$r_-^L=r_-^R$, $r_+^L=r_+^R$ and, consequently, $\la_-^L=\la_-^R$, $\la_+^L=\la_+^R$.
In this case we obtain a new type of structures: contact dispersive shock wave.
For this situation, the parabolas corresponding to $r_-^L=\mathrm{const}$ and $r_-^R=\mathrm{const}$
in Fig.~\ref{Fig7} coincide with each other and a cnoidal DSWs disappear.
Instead, there appears the path connecting the identical left and right states
labeled by the crossing points of two parabolas as is shown in Fig.~\ref{Fig9}.
Such waves can arise only if the boundary points are located on the opposite
sides of the line $u=1/\al$, i.e. in different regions of monotonicity.

\begin{figure}[t] \centering
\includegraphics[width=6cm]{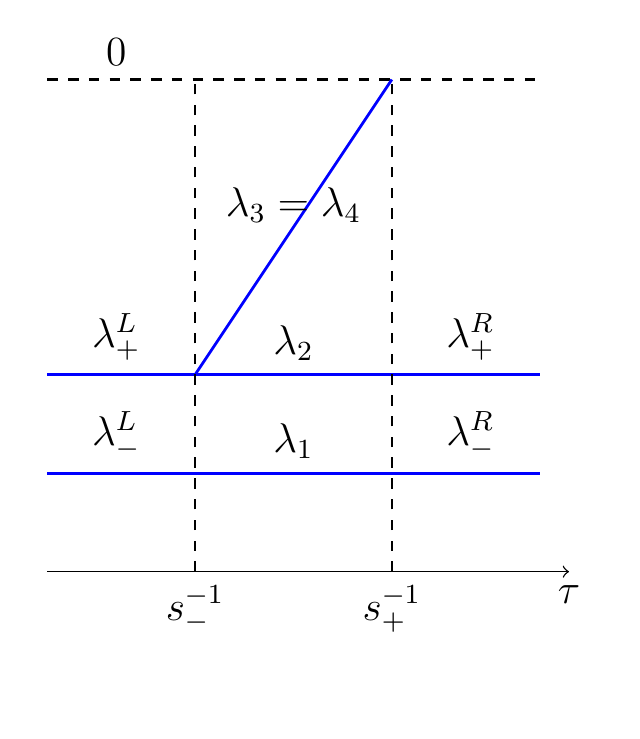}
\caption{Diagram represents evolution of the Riemann invariants as functions of $\tau=t/x$
in the contact DSW solution of the Whitham equations: $r_-^L=r_-^R$ ($\lambda_-^L=\lambda_-^R$),
$r_+^L=r_+^R$ ($\lambda_+^L=\lambda_+^R$).}
\label{Fig10}
\end{figure}

In this situation shown in
Fig.~\ref{Fig10}, the invariants $\lambda_1$ and $\lambda_2$ are
constant within the shock region and they match the boundary conditions
$\lambda_1=\lambda_-^L=\lambda_-^R$, $\lambda_2=\lambda_+^L=\lambda_+^R$, whereas the two other
Riemann invariants remain equal to each other along the shock ($\lambda_3=\lambda_4$) and satisfy the same
Whitham equation $v_3^{-1}(\la_-^L,\la_+^L, \lambda_4(\tau),\lambda_4(\tau)) = v_4^{-1}(\la_-^L,\la_+^L,
\lambda_4(\tau),\lambda_4(\tau))=\tau$. Thus we obtain
\begin{equation}\label{}
    \begin{split}
    & \lambda_1=\lambda_-^L=\lambda_-^R, \quad \lambda_2=\lambda_+^L=\lambda_+^R,\\
    & \frac1{v_4}=\frac{1}{\al}-
        4\lambda_4^2-\frac{((\lambda_+^L)^2-(\lambda_-^L)^2)^2}{(\lambda_+^L)^2+(\lambda_-^L)^2-2\lambda_4^2}=\tau,
    \end{split}
\end{equation}
where the last formula determines the dependence of $\lambda_4$ on $\tau$,
which can be presented in the explicit form
\begin{equation}\label{}
    \begin{split}
    \lambda_4^2(\tau)=&\frac18
     \bigg\{2\left((\lambda^R_-)^2+(\lambda^R_+)^2\right)+\frac{1}{\al}-\tau \\
    & -\bigg[\left(2\left((\lambda^R_-)^2+(\lambda^R_+)^2\right)-\frac{1}{\al}+\tau\right)^2\\
    & + 8\left((\lambda^R_-)^2-(\lambda^R_+)^2\right)^2\bigg]^{1/2}\bigg\}.
    \end{split}
\end{equation}
Here $\tau$ varies within the interval $s_-^{-1}\leq \tau\leq s_+^{-1}$ with
\begin{equation}\label{}
\begin{split}
    s_-^{-1}&=\frac{1}{\al}-(\lambda^R_-)^2-3(\lambda^R_+)^2, \\
    s_+^{-1}&=\frac{1}{\al}-\frac{\left((\lambda^R_-)^2-(\lambda^R_+)^2\right)^2}{(\lambda^R_-)^2+(\lambda^R_+)^2}.
    \end{split}
\end{equation}
The period in this case is given by the formula
\begin{equation}\label{}
    T=\frac{\pi}{2\sqrt{\left(\lambda_4^2-(\lambda_-^L)^2)(\lambda_4^2-(\lambda_+^L)^2\right)}}.
\end{equation}

\begin{figure}[t] \centering
\includegraphics[width=4cm]{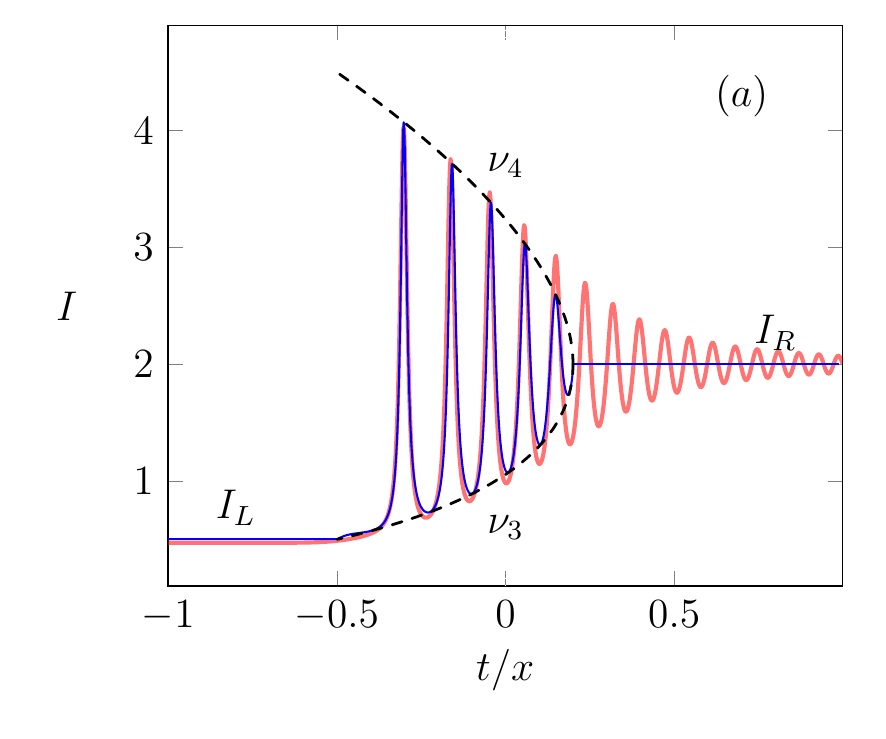}
\includegraphics[width=4cm]{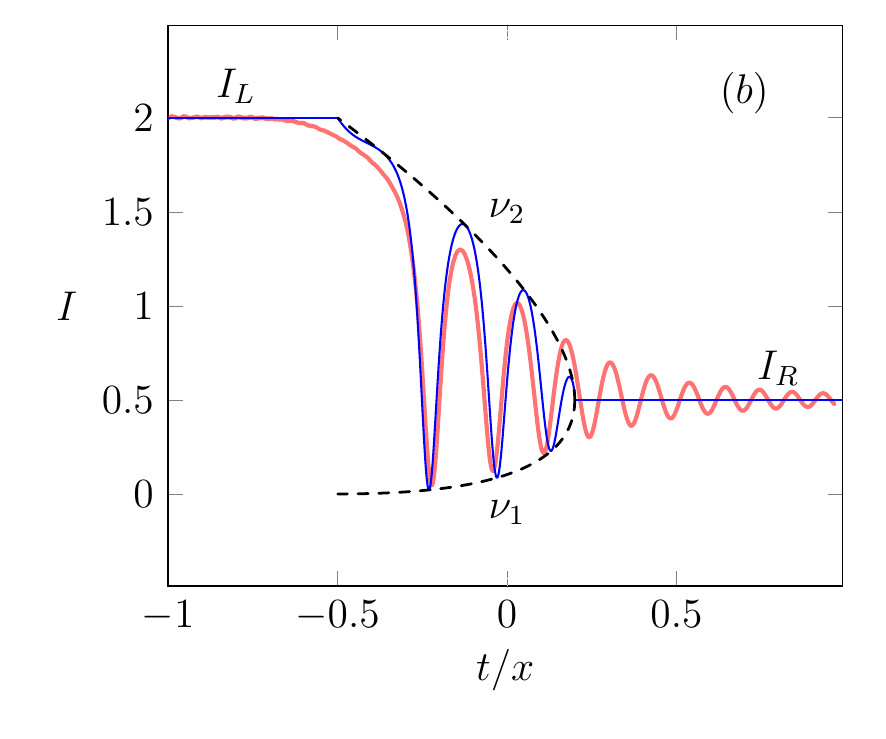}
\caption{Comparison of analytical and numerical solutions of the mNLS equation (\ref{DNLS1})
with contact DSW for two possible choices of directions $P_1\longleftrightarrow P_2$ and
corresponding mappings (\ref{zeros1}) and (\ref{zeros2}).
Here $\alpha=1$ and (a) $I^L=0.5$, $u^L=-0.5$, $I^R=2$, $u^R=2.5$;
(b) $I^L=2$, $u^L=2.5$, $I^R=0.5$, $u^R=-0.5$.
The Riemann invariants are equal to $r_+^L=r_+^R=0.25$ and $r_-^L=r_-^R=-1.75$.
Thin line corresponds to the analytic
solution, thick gray line to numerics, dashed lines show analytical envelopes.}
\label{Fig11}
\end{figure}

As in the case of cnoidal DSWs, the single contact DSW diagram Fig.~\ref{Fig10}
corresponds to two structures due to different mappings
(\ref{zeros1}) or (\ref{zeros2}).
Example of such a structure is shown in Fig.~\ref{Fig11} where
in the first case (a), the path $P_1 \rightarrow P_2$ is realized and formulae (\ref{zeros2}) are used,
and in the second case (b) the opposite path $P_1 \leftarrow P_2$ takes place with
corresponding  formulae (\ref{zeros1}).

\subsection{Combined shocks}

\begin{figure}[t] \centering
\includegraphics[width=8.5cm]{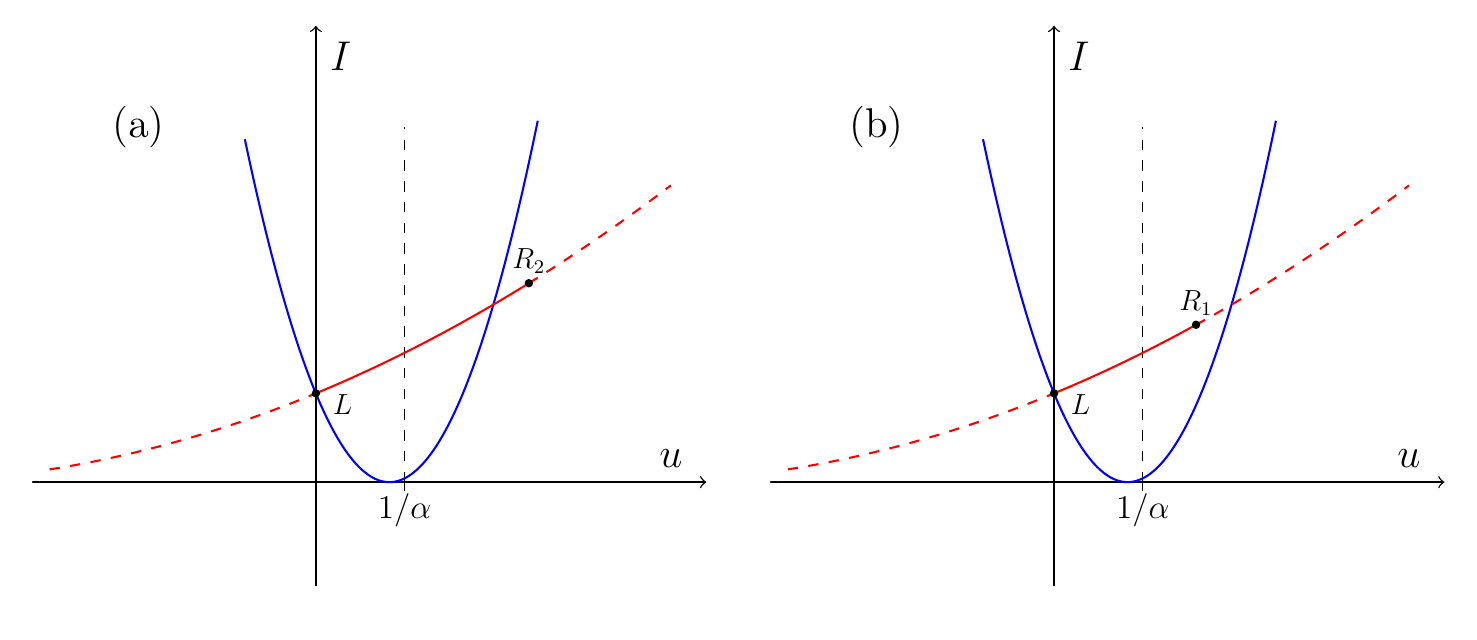}
\caption{Paths in the $(u,I)$-plane associated with two types of combined shocks.
The left and right boundary conditions correspond to points $L$ and $R$ respectively;
they lie on the parabolas along which the dispersionless Riemann invariant $r_-=r_-^L=r_-^R$ ($\la_-=\la_-^L=\la_-^R$) is constant.
One has $r_+^L<r_+^R$ ($\la_+^L<\la_+^R$) in case (a) and $r_+^L>r_+^R$ ($\la_+^L>\la_+^R$) in case (b).}
\label{Fig12}
\end{figure}

\begin{figure}[t] \centering
\includegraphics[width=8cm]{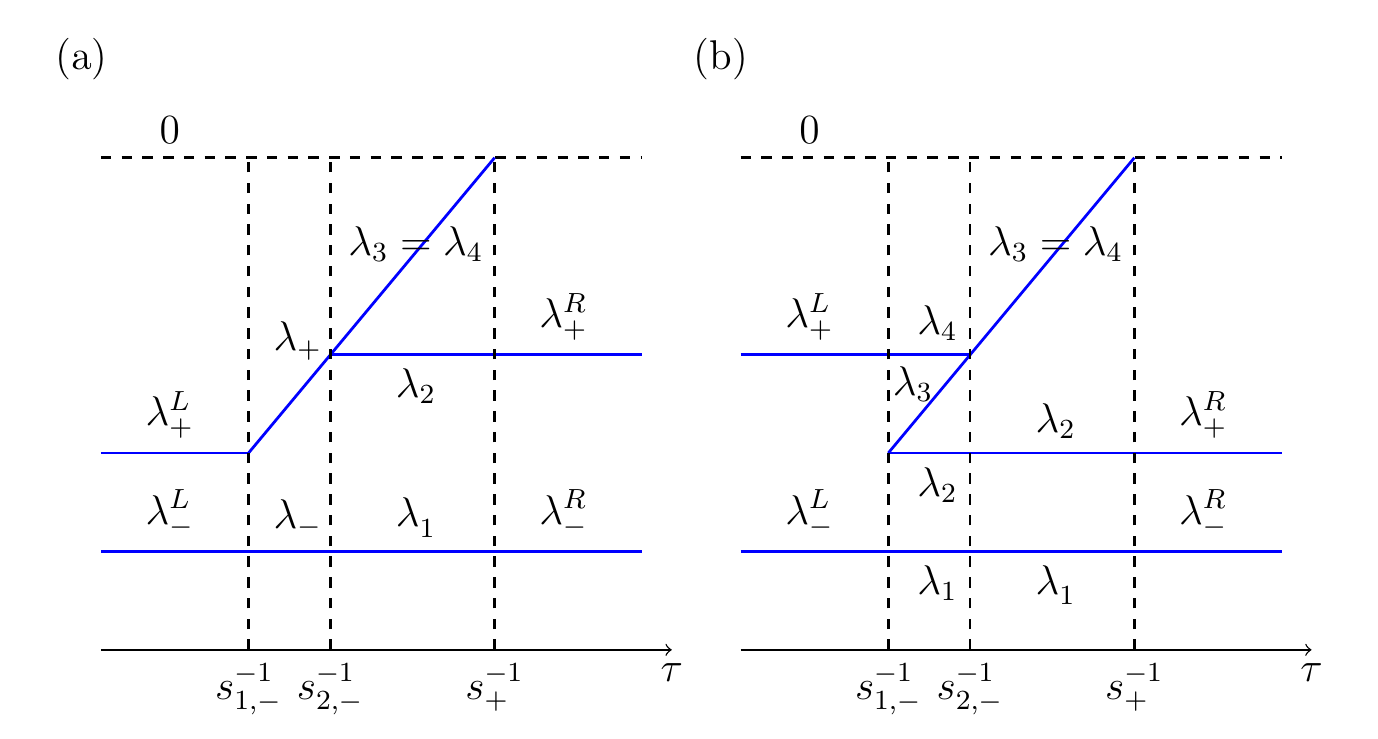}
\caption{Diagram representing the evolution of the Riemann invariants as functions
of $\tau=t/x$ for combined shocks corresponding to the paths in the $(u,I)$-plane shown in Fig.~\ref{Fig12}.
Thin line corresponds to the analytic
solution, thick gray line to numerics, dashed lines show analytical envelopes.}
\label{Fig13}
\end{figure}

Now we turn to the last elementary structure connecting two plateau states and
therefore it can be symbolized by a single path between two points in the $(u,I)$-plane.
This type of paths is illustrated in Fig.~\ref{Fig12} and obviously it is a generalization
of the preceding structure. In this case the boundary points are also located in
different monotonicity regions. One of the Riemann invariants still remains constant
($r_-^L=r_-^R$ or $\la_-^L=\la_-^R$), however the boundary values of the other Riemann invariant are different:
we have $r_+^L<r_+^R$ in case (a) and $r_+^L>r_+^R$ in case (b).
The corresponding $\la$-diagrams are shown in Fig.~\ref{Fig13}.

\begin{figure}[t] \centering
\includegraphics[width=4cm]{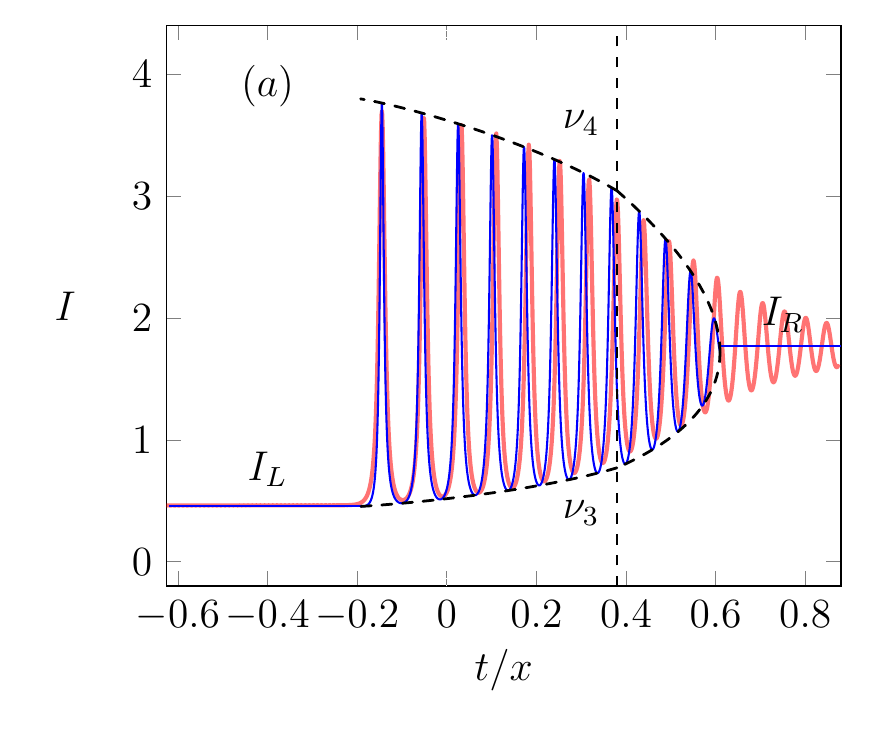}
\includegraphics[width=4cm]{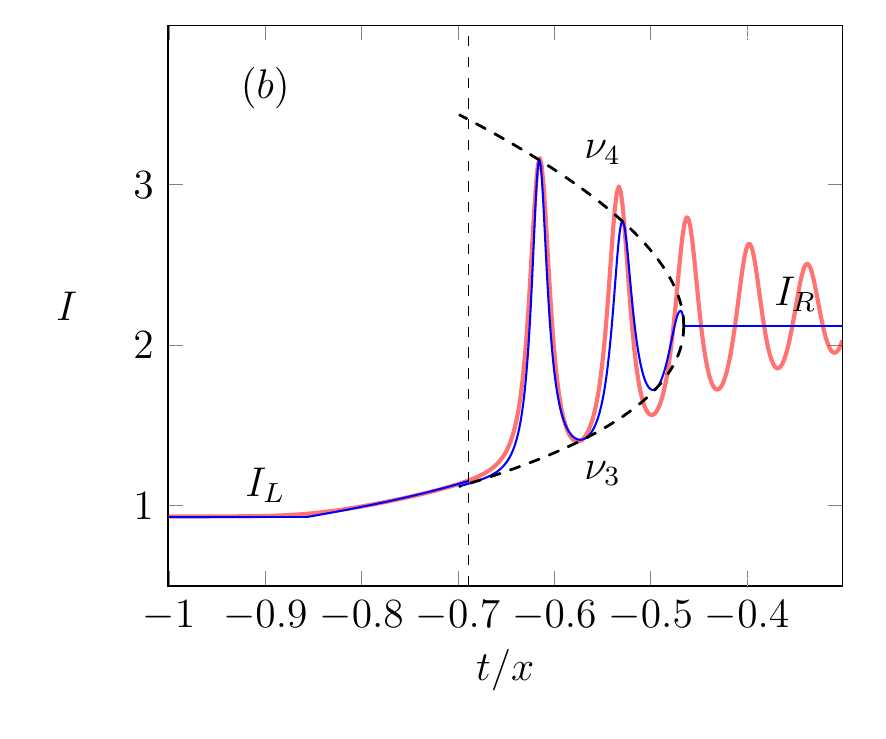}
\caption{Comparison of analytical and numerical solutions of the mNLS equation (\ref{DNLS1})
for combined shocks corresponding to the paths in the $(u,I)$-plane (Fig.~\ref{Fig12})
and to the diagrams of Riemann invariants, Fig.~\ref{Fig13}.
Here $\alpha=1$ and (a) $I^L=0.5$, $u^L=0.2$, $I^R=1.768$, $u^R=2.5$; (b) $I^L=0.94$, $u^L=-0.46$, $I^R=2.12$, $u^R=2$.}
\label{Fig14}
\end{figure}

In the case corresponding to Fig.~\ref{Fig13}(a) the contact dispersive shock wave is attached
at its soliton edge to the rarefaction wave which matches at its left
edge with the left boundary plateau. The velocities of the
characteristic points identified in Fig.~\ref{Fig13}(a)
are expressed in terms of the boundary
Riemann invariants by the formulae
\begin{equation}\label{}
    \begin{split}
    & s_{1,-}^{-1}=\frac1{\al}-\left((\lambda_-^L)^2+3(\lambda_+^L)^2\right), \\
    &   s_{2,-}^{-1}=\frac1{\al}-\left(3(\lambda_+^R)^2+(\lambda_-^R)^2\right),\\
    & s_+^{-1}=\frac{1}{\al}-\frac{((\lambda_+^R)^2-(\lambda_-^R)^2)^2}{(\lambda_+^R)^2+(\lambda_-^R)^2}.
    \end{split}
\end{equation}
The resulting composite wave structure is shown in Fig.~\ref{Fig14}(a)
(thin black line) where it is compared with the numerical solution of the
mNLS equation (thick gray (red) line).

In the case corresponding to Fig.~\ref{Fig13}(b) the trigonometric
DSW is attached at its left edge to the cnoidal dispersion shock wave.
At the left soliton edge the cnoidal wave matches with the left
boundary plateau. The velocities of the characteristic points
identified in Fig.~\ref{Fig13}(b) are given by
\begin{equation}\label{si14}
    \begin{split}
    & s_{1,-}^{-1}=\frac{1}{\al}-\left((\lambda_-^L)^2+2(\lambda_+^R)^2+(\lambda_+^L)^2\right), \\
    & s_{2,-}^{-1}=\frac{1}{\al}-4(\lambda_+^L)^2-
        \frac{((\lambda_+^R)^2-(\lambda_-^R)^2)^2}{(\lambda_+^R)^2+(\lambda_-^R)^2-2(\lambda_+^L)^2},\\
    & s_+^{-1}=\frac{1}{\al}-\frac{((\lambda_+^R)^2-(\lambda_-^R)^2)^2}{(\lambda_+^R)^2+(\lambda_-^R)^2}.
    \end{split}
\end{equation}
The resulting composite wave structures are shown in Fig.~\ref{Fig14}(b)
(blue lines) where they are compared with the numerical solution of the
mNLS equation (red lines).

Now, after description of all elementary wave structures arising in evolution of discontinuities
in the mNLS equation theory, we are in position to formulate the main principles of
classification of all possible wave structures.

\section{Classification of wave patterns}\label{sec6}

Classification of possible structures is very simple in the KdV equation case when any
discontinuity evolves into either rarefaction wave, or cnoidal DSW \cite{gp-1973}.
It becomes more complicated in the NLS equation case \cite{gk-1987,el-1995} and
similar situations as, e.g., for the Kaup-Boussinesq equation \cite{egp-2001,cikp-2017},
where the list consists of eight or ten structures which can be seen after simple enough
inspection of available possibilities and studied one by one. However, the situation
changes drastically when we turn to non-convex dispersive hydrodynamics: even in the case
of unidirectional Gardner (mKdV) equation we get eight different patterns (instead of two
in KdV case) due to appearance of new elements (kinks or trigonometric and combined
dispersive shocks), but these patterns can be labeled by two parameters only and
therefore these possibilities can be charted on a two-dimensional diagram.
In our present case the initial discontinuity (\ref{init}) is parameterized by
four parameters $u^L,I^L,u^R,I^R$, hence the number of possible wave patterns considerably
increases and it is impossible to present them in a two-dimensional chart.
Therefore it seems more effective to formulate the principles according to which
one can predict the wave pattern evolving from a discontinuity with given parameters.
Similar method is used \cite{ivanov-2017} in classification of wave patterns evolving
from initial discontinuities according to the Landau-Lifshitz equation for easy-plane
magnetics or polarization waves in two-component Bose-Einstein condensate.

\begin{figure}[t] \centering
\includegraphics[width=7cm]{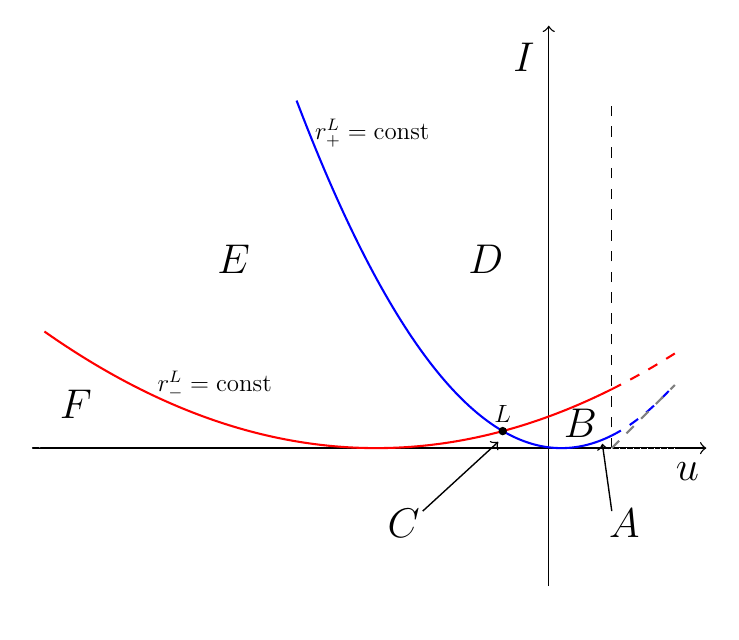}
\caption{Domains in the left monotonicity region of the ($u,I$)-plane
corresponding to different wave structures. }
\label{Fig15}
\end{figure}

It is convenient to begin with the consideration of the classification problem from the case
when both boundary points lie on one side of the line $u=1/\al$ separating two
monotonicity regions in the $(u,I)$-plane.
At first we shall consider situation when the boundary points lie in the left monotonicity region.
We show in Fig.~\ref{Fig15} the two parabolas corresponding to the constant
dispersionless Riemann invariants $r_{\pm}^L$ related with the left boundary state.
Evidently, they cross at some point $L(u^L,I^L)$ where $r_-^L=r_+^L$. These two
parabolas cut the left monotonicity region into six domains labeled
by the symbols $A,B,\ldots,F$.  Depending on the domain, in which the point $R$ with
coordinates $(u^R,I^R)$, representing the right boundary condition, is located,
one gets one of the six following possible orderings of the left and
right Riemann invariants:
\begin{equation}\label{RiemannInequalities}
    \begin{split}
    & \mbox{A}: \quad \la_-^R < \la_+^R < \la_-^L < \la_+^L, \\
     &  \mbox{B}: \quad \la_-^R < \la_-^L < \la_+^R < \la_+^L, \\
    & \mbox{C}: \quad \la_-^L < \la_-^R < \la_+^R < \la_+^L, \\
     &   \mbox{D}: \quad \la_-^R < \la_-^L < \la_+^L < \la_+^R, \\
    & \mbox{E}: \quad \la_-^L < \la_-^R < \la_+^L < \la_+^R, \\
     &   \mbox{F}: \quad \la_-^L < \la_+^L < \la_-^R < \la_+^R.
  \end{split}
\end{equation}
All these six domains and corresponding orderings yield six possible wave
structures evolving from initial discontinuities. Let us consider briefly each of them.

\begin{figure}
\centering
\includegraphics[width=4cm]{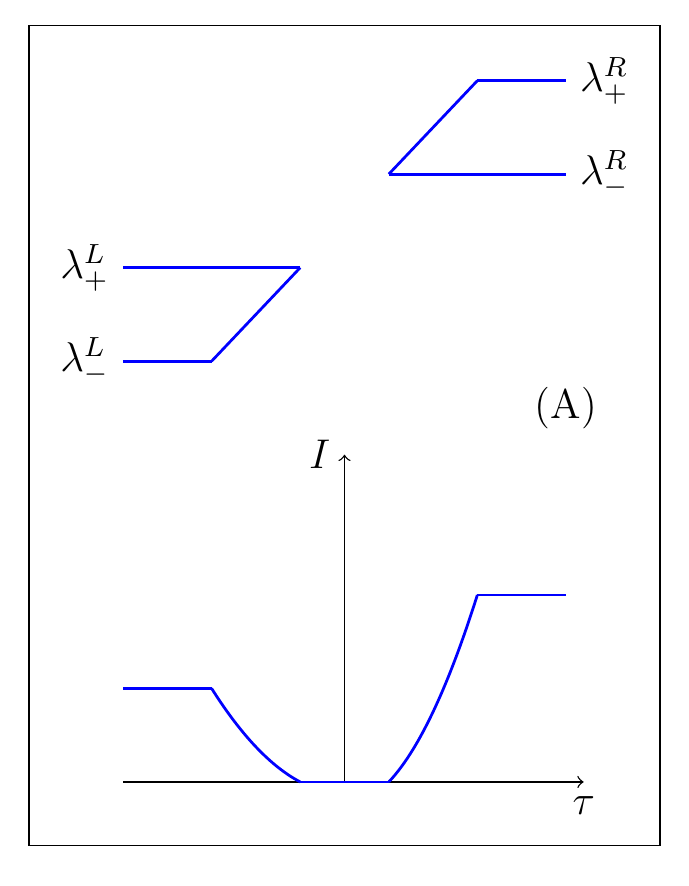}
\includegraphics[width=4cm]{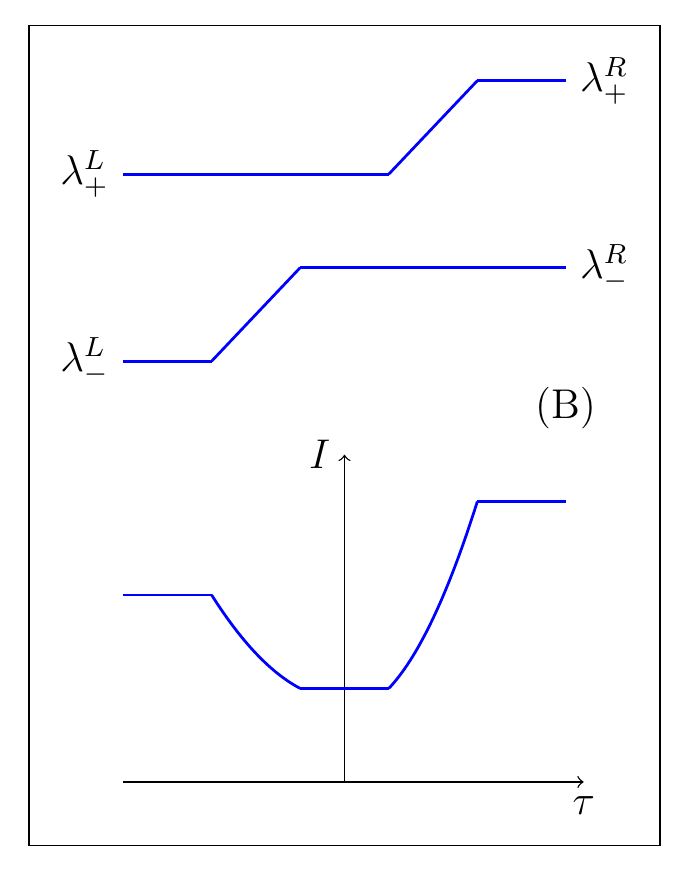}\\
\includegraphics[width=4cm]{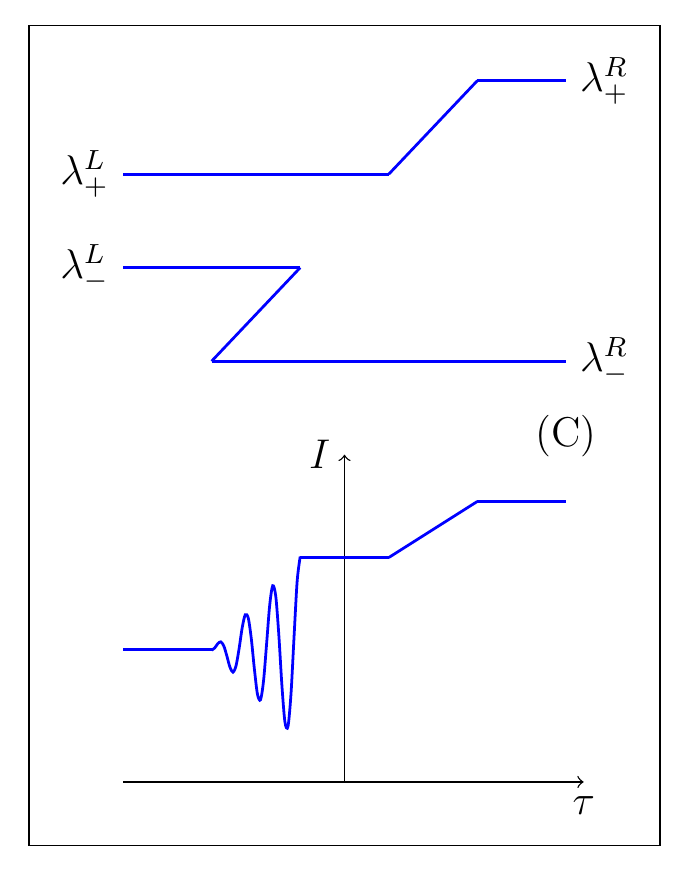}
\includegraphics[width=4cm]{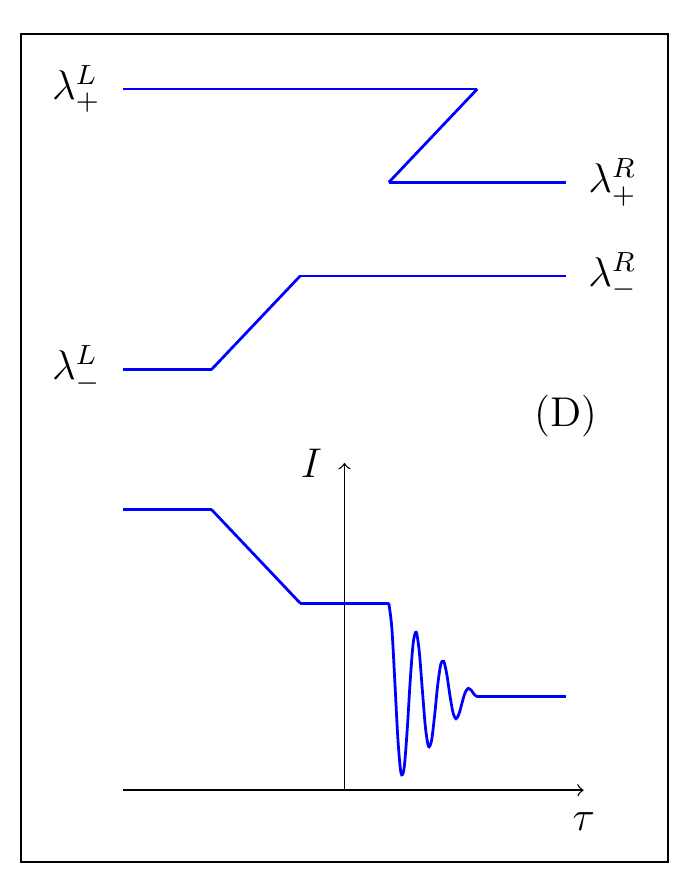}\\
\includegraphics[width=4cm]{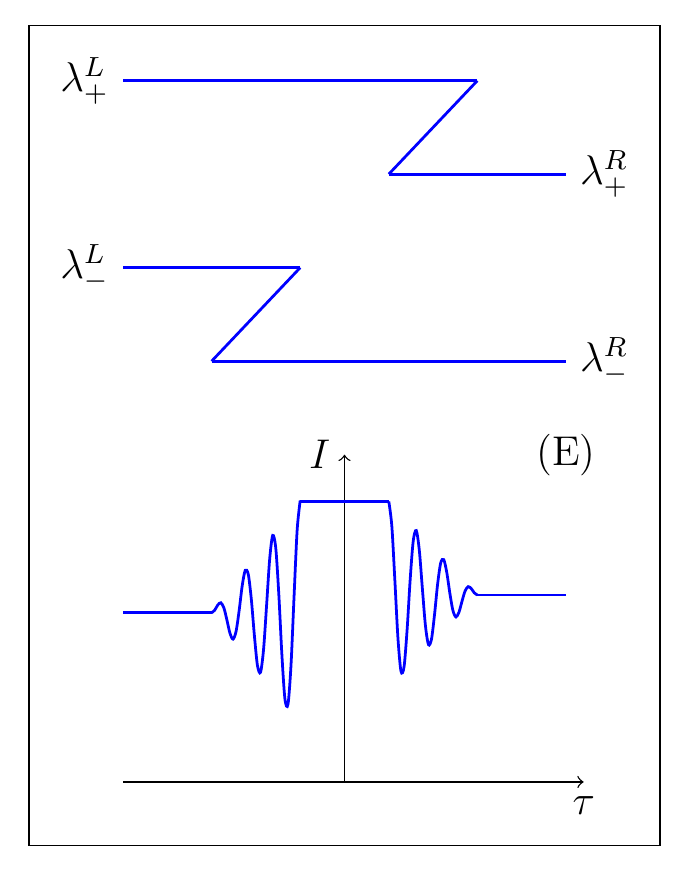}
\includegraphics[width=4cm]{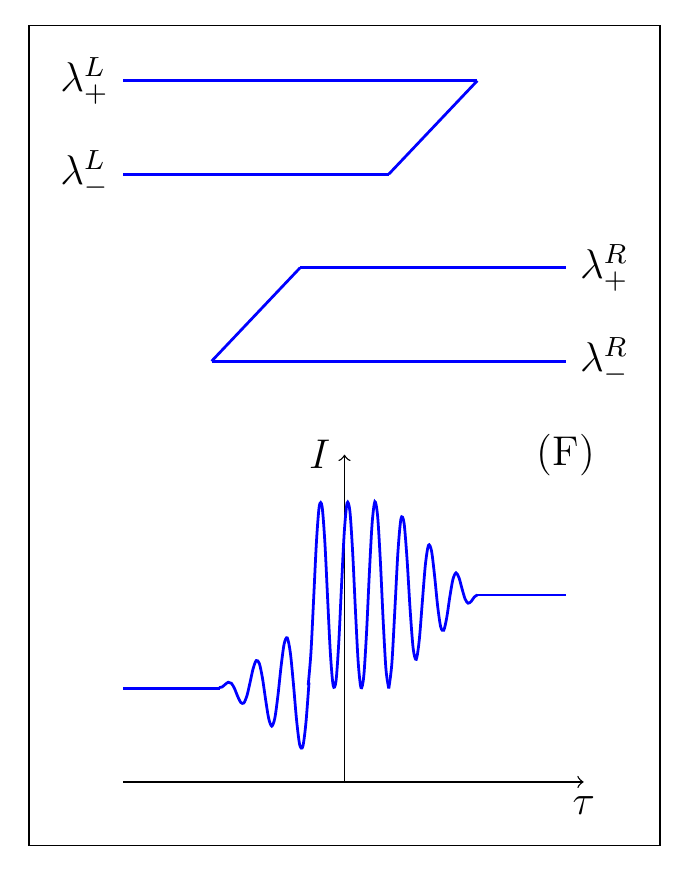}
\caption{Sketches of behavior of the Riemann invariants and of the corresponding wave
structures for six possible choices of the boundary conditions. }
\label{Fig16}
\end{figure}

$\bullet$ In case (A) two rarefaction waves are combined into a single wave structure where
they are separated by an empty region.
This means that two light fluids flow in opposite directions with velocities so large that the
rarefaction waves are not able to fill in an empty region between them.
Evolution of Riemann invariants and sketch of wave structure
are shown in Fig.~\ref{Fig16}(A).

$\bullet$ In case (B) two rarefaction waves are connected by a plateau whose
parameters are determined by the dispersionless Riemann invariants
$r_{\pm}^P$ equal to $r_-^P=r_-^R$ and $r_+^P=r_+^L$.
Here rarefaction waves are able now to provide enough flux of the light fluid to create
a plateau in the region between them (see Fig.~\ref{Fig16}(B)).

$\bullet$ In case (C) we obtain a dispersive shock wave on the left,
a rarefaction wave on the right and a plateau in
between are produced (see Fig.~\ref{Fig16}(C)).

$\bullet$ In case (D) we get the same situation as in the case (C),
but now the dispersive shock wave and rarefaction wave exchange their places
(see Fig.~\ref{Fig16}(D)).

$\bullet$ In case (E) two DSWs are produced with a plateau between them.
Here we have a collision of two light fluids (see Fig.~\ref{Fig16}(E)).

$\bullet$ In case (F) the plateau observed in the case (E) disappears.
It is replaced by a nonlinear wave which
can be presented as a non-modulated cnoidal wave (see Fig.~\ref{Fig16}(F)).

The possible structures for this part of the $(u,I)$-plane
coincide qualitatively with the patterns found in similar classification problem
for the nonlinear Schr\"odinger equation \cite{el-1995}.
It is clear that as $\al$ tends to zero, the mNLS equation transforms to the nonlinear Schr\"odinger equation.
Then the line $u=1/\al$ goes to infinity and therefore there remains only the left monotonicity region.

\begin{figure}[t] \centering
\includegraphics[width=8cm]{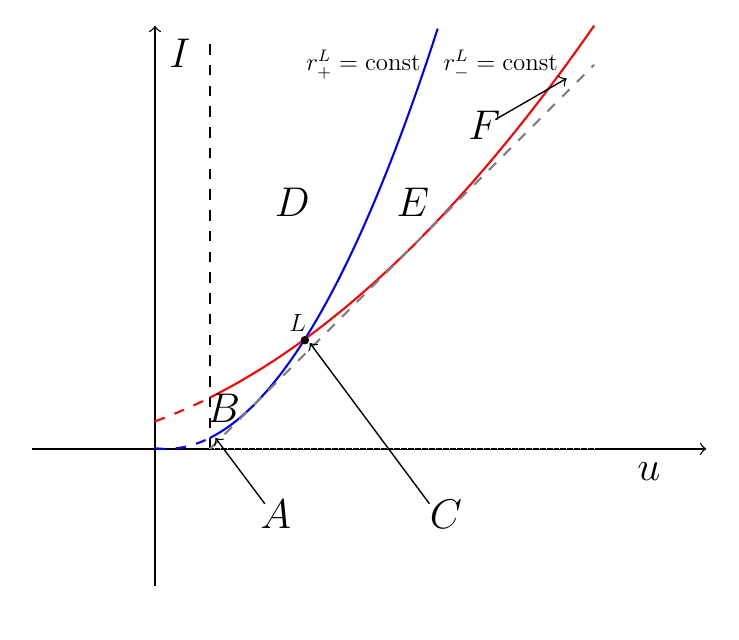}
\caption{Domains in the ($u,I$)-plane on the right side of the line $u=1/\al$ corresponding to different
structures. }
\label{Fig17}
\end{figure}

Now we turn to consideration of the classification problem for the case
when both boundary points lie to the right of the line $u=1/\al$.
This situation is shown in Fig.~\ref{Fig16}.
We see that the parabolas divide again this right monotonicity region into
six domains. For this case the Riemann invariants can have
the same orderings (\ref{RiemannInequalities}) as in the previous case.
Depending on the location of the right boundary point in a certain domain, the corresponding
wave structure will be formed. For all cases
these structures coincide with those for the previous case.

At last, we have to investigate the situation when the boundary points lie on different
sides of the line $u=1/\al$, that is in different monotonicity regions.
As we have seen in the previous section, in this case new complex structures consisting
of contact dispersive shock waves or combined shocks appear.
Since the total number of possible wave patterns
is very large, we shall not list all of them here but
rather illustrate the general principles of their classification.

\begin{figure}[t] \centering
\includegraphics[width=8cm]{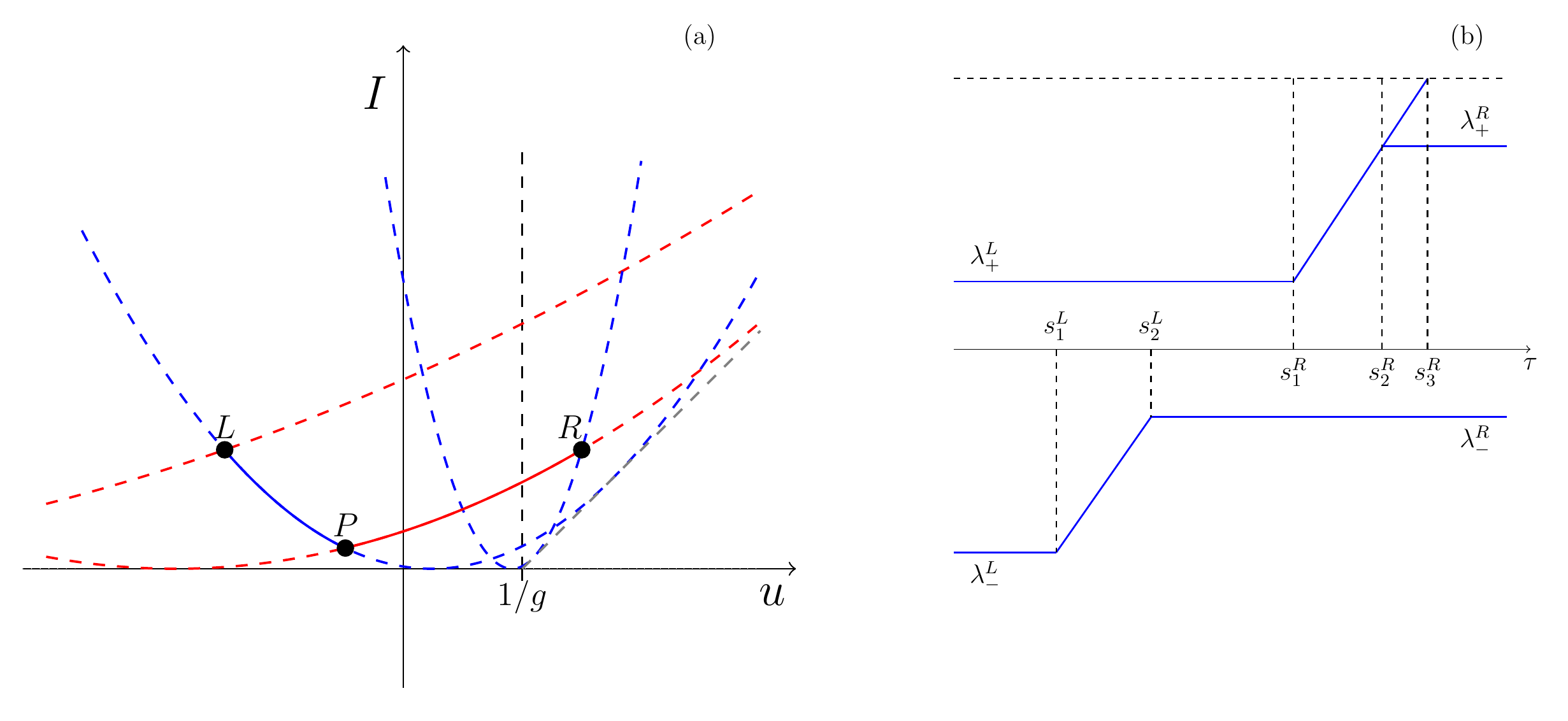}
\caption{(a) The branches of the parabola corresponding to the path between the left and right points.
(b) The corresponding diagram for the Riemann invariants.}
\label{Fig18}
\end{figure}

For given boundary parameters, we can construct the parabolas corresponding to
constant Riemann invariants $r_{\pm}^{L,R}$: each left or right pair of these
parabolas crosses at the point $L$ or $R$ representing the left or right
boundary state's plateau. Our task is to construct the path joining these two
points, then this path will represent the arising wave structure. We already know the answer
for the case when the left and right points lie on the same parabola, see, e.g.,
Fig.~\ref{Fig12}. If this is not the case and the right point $R$ lies, say,
below the parabola $r_-^L=\mathrm{const}$, see Fig.~\ref{Fig18}(a), then we can
reach $R$ by means of more complicated path consisting of two arcs of parabolas
joined at the point $P$. Evidently, this point $P$ represents the plateau between two waves
represented by the arcs. At the same time, each arc corresponds to a wave structure
discussed in the preceding section. In fact, there are
two paths with a single intersection point that join the left and right boundary points,
and one can easily see another path made of dashed lines in Fig.~\ref{Fig18}(a). We choose the
physically relevant path by imposing the condition that velocities of edges of all
regions must increase from left to right.
Having constructed a path from the left boundary point to the right one,
it is easy to draw the corresponding $\la$-diagram.
To construct the wave structure, we use
the formulae connecting the zeros $\nu_i$ of the resolvent with the Riemann
invariants $\lambda_i$ and expressions for the solutions parameterized by $\nu_i$.
This solves the problem of construction of
the wave structure evolving from the initial discontinuity with given boundary conditions.

\begin{figure}[t] \centering
\includegraphics[width=8cm]{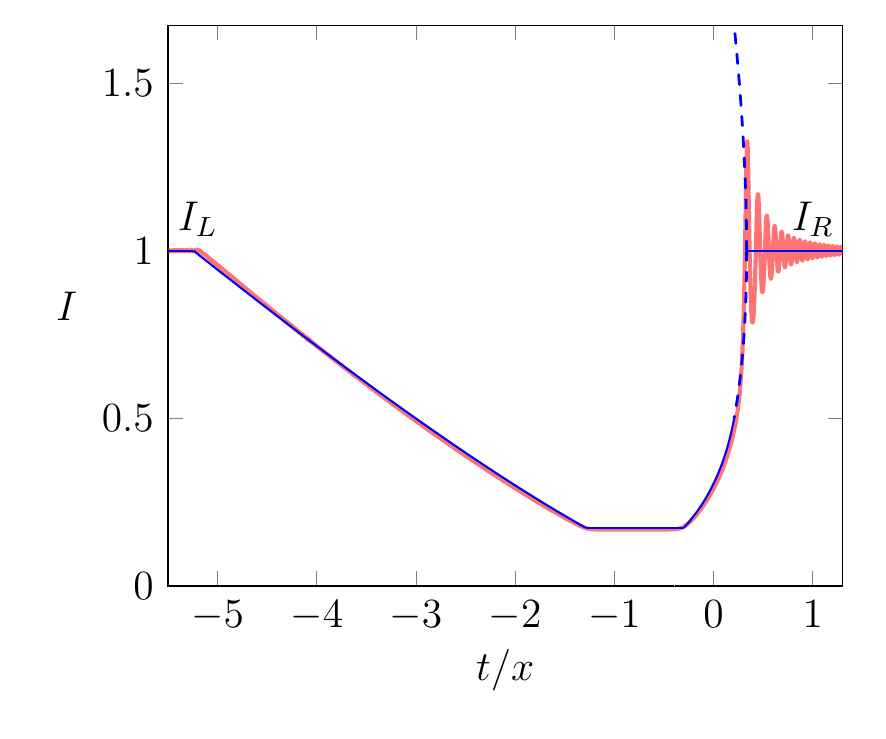}
\caption{The wave structure corresponding to the conditions $I^L=I^R=1$, $u^L=-u^R=-1$
at the initial discontinuity. It is described by the path shown in Fig.~\ref{Fig18}(a)
and $\la$-diagram shown in Fig.~\ref{Fig18}(b). Here $\alpha=1$.
Thin line corresponds to the analytic
solution, thick gray line to numerics.}
\label{Fig19}
\end{figure}
For example, let us consider the case $I^L=I^R=1$, $u^L=-u^R=-1.5$ which corresponds to
Fig.~\ref{Fig18}(a). We see that the branch of the parabola with $r_-^R=\mathrm{const}$
crosses the line $u=1/\al$.
Taking into account that the left wave
corresponds to the continuation of $r_+^L=\mathrm{const}$ and the
right wave to the continuation of $r_-^R=\mathrm{const}$, we arrive at
the diagram shown in Fig.~\ref{Fig18}(b). Consequently,
at the left edge we have a rarefaction wave and at the right edge
the combination of a trigonometric shock
with a rarefaction wave. Between these waves we get a plateau
characterized by the Riemann
invariants $r_-^P=r_-^R$ and $r_+^P=r_+^L$. This plateau is represented by
a single point $P$ in Fig.~\ref{Fig18}(a).
The wave structure can be obtained by substitution of solution of the Whitham
equation into expressions for wave oscillations. As we see,
our analytical results agree very well with numerical calculations
shown in Fig.~\ref{Fig19}.

\begin{figure}[t] \centering
\includegraphics[width=8cm]{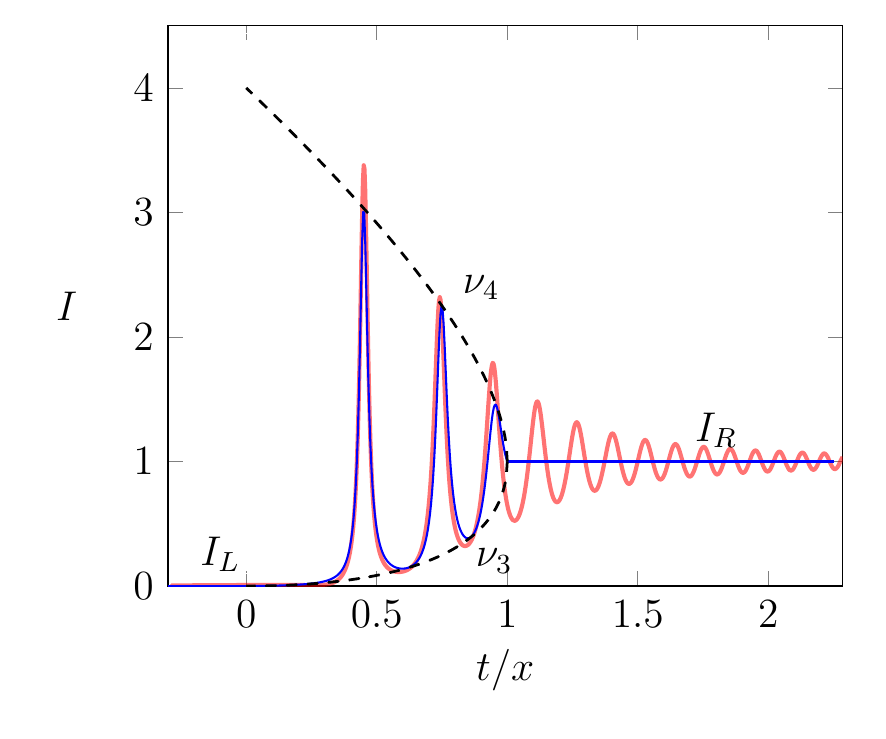}
\caption{Contact dispersive shock for ``expansion into vacuum'' type of the initial discontinuity.
Thin line corresponds to the analytic
solution, thick gray line to numerics, dashed lines show analytical envelopes.}
\label{Fig20}
\end{figure}

Another instructive example describes the situation when we get
``expansion into vacuum'' wave pattern with formation of a contact shock wave.
Such a situation is impossible in the NLS theory \cite{el-1995} where
expansion into vacuum leads always to formation of a rarefaction wave.
However, the mNLS equation (\ref{DNLS1}) differs drastically in this
respect from the NLS equation case.
In the problem of evolution of the initial discontinuity, if velocity and
intensity on the left boundary are equal to zero ($u_L=0$, $I_L=0$),
then the Riemann invariants also vanish at this boundary ($r_-^L=r_+^L=0$).
In spite of that, they can form a contact shock in transition to the right
boundary if the dispersionless Riemann invariants
on the right boundary are also equal zero ($r_-^R=r_+^R=0$).
From equations (\ref{IandUbyR}) we find that there are two possibilities for that:
either $u^R=0$, $I^R=0$ or $u^R=2$, $I^R=1$.
The first trivial option refers to the absence of light in the waveguide
and therefore it is not of any interest. The second option corresponds precisely to
the case of the formation of the contact shock wave.
Fig.~\ref{Fig20} shows such a structure with comparison of the numerical
solution with the analytical one.
This comparison shows that the analytic Whitham theory agrees with numerics very well.
The corresponding diagram of the Riemann invariants qualitatively coincides with the diagram in Fig.~\ref{Fig10}
with one difference: the Riemann invariant $r_-$ coincides with $r_+$.
Therefore the dispersive Riemann invariants $\lambda_1$ and $\lambda_2$ also coincide with each other.

\section{Conclusion}\label{sec7}

In this paper, we have developed the Whitham method of modulations for propagation of long enough
pulses in fibers with account of steepening effects. The theory is applied to the problem of
classification of wave patterns evolving from given discontinuity in the initial data.
Because of non-convex behavior of nonlinear velocities in this case, previously known methods
of solving such kind of problems should be modified with inclusion of new types of elementary
wave structures, such as `contact dispersive shocks'. Evolution of these structures is described by the
degenerate limits of the Whitham modulation equations. In the resulting scheme, one solution of the
Whitham equations corresponds to two different wave patterns, and this correspondence is provided by
a two-valued mapping of Riemann invariants to physical modulation parameters. In this respect,
situation is similar to that of modified KdV case already discussed in Ref.~\cite{kamch-2012},
but here the system with two-directional propagation of waves is considered, and
one can compare this with transition from the KdV equation case \cite{gp-1973} to NLS equation
case \cite{el-1995}.  The resulting set of possible wave patterns is very rich and we have
developed a graphical method for determining which structure will evolve from given initial data.
The method is quite flexible and it was also applied to another system with non-convex
hydrodynamics---Landau-Lifshitz equation for dynamics of magnetics with uniaxial easy-plane
anisotropy \cite{ivanov-2017}.

In principle, one may hope that the results found here can be observed experimentally in systems similar
to that used in the recent experiment \cite{xu-2017}. However, one should keep in mind that in standard
fibers the Raman effect is typically much stronger than the self-steepening effect (see, e.g.,
Ref.~\cite{ka-2003}). Fortunately, the manifestations of these two effects are quite different and therefore
they can be identified separately. As was shown in this paper, the main new effect of the self-steepening
term is formation of combined shocks caused by the non-convex properties of the nonlinearity, whereas the
Raman effect leads to formation of stationary shocks with finite length (see, e.g.,
\cite{kivshar-90, ah-92,km-93,wfr-98}). In the limit of long-time evolution, the combined action of both
effects must lead to formation of combined stationary shocks different from shocks predicted by the theory
which takes into account the Raman effect only. Qualitatively, these shocks must look similar to
those described here. The quantitative theory of this new type of combined
shocks can be developed in framework of the presented here approach, however this task is definitely
beyond the present paper.

Another possibility of observation of predicted here effects is related with the use of photonic crystal
waveguides which are free from the Raman scattering, as it was observed experimentally in
Ref.~\cite{colman-12}, and the waveguides can be engineered in such a way that the self-steepening parameter
is considerably increased \cite{{zzkf-13}}.

Thus, the presented here theory, on one side, predicts some new phenomena which can be observed
experimentally and, on the other side, it forms the basis for development of more complete theories
which take into account other effects.

\section*{Acknowledgments}

We are grateful to M.~Conforti, T.~Congy, A.~Kudlinski, A.~Mussot and N.~Pavloff for useful
discussions at the initial stage of this work.

\end{document}